\documentclass[11pt,twocolumn]{article}

\usepackage{amsmath,amsfonts,amssymb,bbm,enumitem,url,color,graphicx,amsthm,rotating,authblk,soul,siunitx,tabulary,booktabs}
\usepackage[font=footnotesize,labelfont=bf]{caption}
\usepackage[a4paper]{geometry}
\usepackage[round]{natbib}

\newcommand{\blue}{}

\begin{document}

\twocolumn[	\begin{@twocolumnfalse}
\begin{center}
	
	{\bf \Large A systematic comparison of tropical waves over northern Africa. Part I: Influence on rainfall}
	
	\bigskip
	
	\medskip
	
	{\bf Andreas Schlueter\textsuperscript{*1}, Andreas H.~Fink\textsuperscript{1}, Peter Knippertz\textsuperscript{1}, Peter Vogel\textsuperscript{1,2}}
	
	\textsuperscript{1} Institute of Meteorology and Climate Research, Karlsruhe Institute of Technology, Germany \\
	\textsuperscript{2} Institute for Stochastics, Karlsruhe Institute of Technology, Germany
	
	\textsuperscript{*}\url{andreas.schlueter@kit.edu} 
	
	\bigskip
	
	\textbf{This work has been submitted to Journal of Climate.\\Copyright in this work may be transferred without further notice.}
	
	\bigskip

\end{center}

\abstract{
	Low-latitude rainfall variability on the daily to intraseasonal timescale is often related to tropical waves, including convectively coupled equatorial waves, the Madden-Julian Oscillation (MJO), and tropical disturbances. Despite the importance of rainfall variability for vulnerable societies in tropical Africa, the relative influence of tropical waves for this region is largely unknown. This article presents the first systematic comparison of the impact of six wave types on precipitation over northern tropical Africa during the transition and full monsoon seasons, using two satellite products and a dense rain gauge network.\\
	\indent \blue{Composites of rainfall anomalies in the different datasets show} comparable modulation intensities in the West Sahel and at the Guinea Coast, varying from less than 2 to above \SI{7}{\mm\per\day} depending on the wave type. African Easterly Waves (AEWs) and Kelvin waves dominate the 3-hourly to daily timescale and explain 10--\SI{30}{\percent} locally. On longer timescales (7--20d), only the MJO and equatorial Rossby (ER) waves remain as modulating factors and explain about up to one third of rainfall variability. Eastward inertio-gravity waves and mixed Rossby-gravity (MRG) waves are comparatively unimportant. \blue{An analysis of wave superposition shows that low-frequency waves (MJO, ER) in their wet phase amplify the activity of high-frequency waves (TD, MRG) and suppress them in the dry phase.} The results stress that more attention should be paid to tropical waves when forecasting rainfall over northern tropical Africa. \\
	\bigskip
}

\end{@twocolumnfalse}]


\section{Introduction}  \label{sec:Introduction}

Rainfall variability substantially affects societies in northern tropical Africa \citep{Sultan.2005}. More than \SI{96}{\percent} of cultivated land in Sub-Saharan Africa is rainfed \citep{FAO.2016}. Despite this, major operational global weather prediction models still fail to deliver skillful short-range precipitation forecasts over this region \citep{Vogel.2018}. This corroborates the need for an improved understanding of the underlying processes involved in the generation of precipitation and their representation in numerical weather prediction (NWP) models over northern tropical Africa.

Equatorial waves are a potential source of predictability in the tropics, as they can be considered preferred eigenmodes of the tropical atmosphere. These waves can couple with deep convection and subsequently modulate rainfall on the synoptic to subseasonal timescale throughout the tropics \citep{Wheeler.1999}. Equatorial waves interacting with precipitation are thus called convectively coupled equatorial waves (CCEWs). Several different types of CCEWs have been identified, which differ in their wavelength and period as well as in their accompanying influence on the dynamic and thermodynamic environment: The solutions of the shallow-water equations, a theory describing wave propagation trapped at the equator, include the equatorial Rossby (ER) wave, the mixed-Rossby-gravity (MRG) wave, the Kelvin wave as well as eastward and westward propagating inertio-gravity (EIG and WIG) waves \citep{Matsuno.1966}. Two major other wave types have been observed in the tropical belt that are not obtained from the shallow-water equations: The Madden-Julian Oscillation (MJO, \citealt{Madden.1971}) and westward traveling tropical disturbances (TD\blue{) including easterly waves (}\citealt{Riehl.1945}). All waves these collectively termed tropical waves in this paper. The spatiotemporal scales range from planetary and 30--90 days in case of the MJO to synoptic and 1--3 days in case of inertio-gravity waves.

Rainfall over northern tropical Africa exhibits substantial variability at the synoptic to intraseasonal timescales. Three different regimes have been identified: the synoptic timescale, with oscillatory signals of periods below 10 days, the short intraseasonal timescale with periods from 10-25 days, and the long intraseasonal timescale with periods from 25-60 days \citep{Janicot.2011}. On the short intraseasonal timescale, two main modes have been \blue{documented}: the 'quasi-biweekly zonal dipole' (QBZD, \citealt{Mounier.2008}) and the 'Sahel' mode \citep{Sultan.2003}. Besides radiative processes, the QBZD has been associated with the presence of Kelvin waves, whereas ER appear to contribute to the 'Sahel' mode \citep{Janicot.2010}. \blue{Finally, the variability of the Saharan Heat Low also modulates rainfall at the timescale of 10--25 days \citep{Chauvin.2010,Roehrig.2011}.}

The dominant influence of African Easterly Waves (AEWs) on precipitation over West Africa has been known several decades \citep{Reed.1977}. AEWs modulate precipitation on the timescale of 2--6 days. About one third of the total variance in deep convection can be explained by 2--6 day filtered disturbances \citep{Dickinson.2000,Mekonnen.2006,Lavaysse.2006,Skinner.2013}. According to \citet{Fink.2003}, more than \SI{60}{\percent} of squall lines in West Africa, which are responsible for the vast majority of rainfall in this region, are associated with AEWs. \citet{Roundy.2004} show that TDs, which are observed in the entire tropics, correspond to African Easterly Waves (AEW) over northern tropical Africa. Following \citet{Roundy.2004}, the term TD will refer to AEWs throughout this paper.

Kelvin waves modulate rainfall on the synoptic to intraseasonal timescale. During the boreal spring, rainfall over central Africa is significantly influenced by Kelvin waves  \citep{Nguyen.2008,Laing.2011,Sinclaire.2015}. \citet{Mekonnen.2008} demonstrate that Kelvin waves double the observed rainfall amount over tropical Africa during the full monsoon. Over West Africa, Kelvin waves explain 10 to \SI{15}{\percent} of convective variability \citep{Mekonnen.2016}. Kelvin waves are not isolated features, instead they also interact with AEWs or trigger them \citep{Mekonnen.2008,Ventrice.2013,Mekonnen.2016}.

On the intraseasonal timescale, the influence of the MJO on precipitation over northern tropical Africa has been discussed by several authors (e.g. \citealt{Matthews.2004,Janicot.2009,Lavender.2009,Ventrice.2011,Alaka.2017}). The MJO significantly modulates the dry spell frequency in the West Sahel as station data suggest \citep{Pohl.2009}. \citet{Gu.2009} and \citet{Pohl.2009} document a more pronounced modulation of precipitation along the Guinean belt. During the spring season, the MJO was found to modulate precipitation in central Africa \citep{Berhane.2015}. 

ER waves mainly modulate precipitation on the subseasonal timescale. In a detailed study of ER over West Africa, \citet{Janicot.2010} distinguish two different types of ER waves. The imprint of the first mode on the 30 to 100 day timescale is linked to the MJO and modulates convection over West and Central Africa. On the timescale of 10 to 30 days, ER waves also modulate precipitation in the Sahelian band. The influence of the \blue{slightly tilted poleward and eastward} second mode can be seen reaching far into subtropics and suggests a link to extratropical circulation. The slightly eastwards tilted precipitation pattern resemble tropical plumes regularly observed during the dry winter season \citep{Knippertz.2005,Frohlich.2013}. 

In the literature, little attention has been paid to MRG waves over this region. Results by \citet{Skinner.2013} indicate that MRG contribute less to rainfall variability than TD, Kelvin, and ER waves. They are most frequently observed in the central and western Pacific region, where they significantly affect precipitation \citep{Takayabu.1993,Holder.2008,Kiladis.2016}. MRG waves occur in a wave domain next to TD. Over the Pacific Ocean, the transition of MRG to off-equatorial TDs has been documented \citep{Takayabu.1993,Zhou.2007}. Similarly, "hybrid" AEW/MRG have been documented over Africa \citep{Cheng.2018}. However, the involved mechanisms are still not fully understood. More research is needed to clarify how MRG waves and AEW interact and to integrate disturbances that do not follow either of their characteristics, as mentioned by \citet{Knippertz.2017}.

Only few studies have focused on EIG \blue{and WIG} waves so far. Dry EIG waves in the mesosphere and stratosphere have been recorded by \citet{Mayr.2003,Mayr.2004}, \citet{Tindall.2006}, and \citet{Tindall.2006b}. Some evidence suggests that EIG can be also found in the troposphere with main influence on upper levels \citep{Yang.2003,Yang.2007,Kiladis.2009}, \blue{although they are not completely separable from MRG waves \citep{Kiladis.2016,Dias.2016}}. \blue{WIG waves were identified to be associated with squall lines which are the main cause for precipitation in the Sahel region \citep{Tulich.2012}.} Whether EIG \blue{and WIG} waves also influence rainfall variability over northern tropical Africa is largely unknown. 

Despite the importance of the different types of tropical waves on the variability of rainfall on synoptic to intraseasonal timescales over northern tropical Africa, as of today, a systematic investigation of the relative influence of the different waves on rainfall variability has not been performed. Several questions concerning the modulation of rainfall by the different waves remain open.  The quantitative influence of tropical waves and the relative contribution to rainfall variability at the different timescales are not well documented and a consistent analysis for all waves using a unified method is pending. Operational forecasters in the region mostly rely on their experience or subjective view when assessing tropical waves because little is known about the relative influence of the different waves at the different timescales, as it has not yet been quantified systematically (B. Lamptey, personal communication, 2017). The aim of the present study is therefore to close this gap and compare the influence of the major types of tropical waves for the monsoon system of Africa using satellite products and in-situ measurements, and one consistent method for all waves. The forecast verification study of \citet{Vogel.2018} has stressed the need to use data from rain gauge networks due to large discrepancies between satellite and rain gauge accumulations in this region at daily time scales. Thus, the present study also uses an extensive database of African rain gauges.

This article will present a comparative study of the influence of tropical waves on precipitation over northern tropical Africa. As the region receives its main rainfall during the 'full monsoon' and the months before and after the onset in the so-called 'transition season', this paper will focus on these two seasons. This study follows in parts the work of \citet{vanderLinden.2016} who performed a similar systematic study on the modulation of precipitation by three tropical waves over Vietnam. The following research questions will be addressed:
\begin{itemize}  
	\item Where, in which season, and how strongly do tropical waves contribute to rainfall variability over northern tropical Africa?
	\item What is their effect on spatial precipitation patterns and observed rainfall amounts?
	\item What is their relative contribution to rainfall variability on different timescales?
	\item Do different types of tropical waves interact when they superimpose?
\end{itemize}
After the description of the used data and filtering methods in section~\ref{sec:Methods}, answers to these four questions are presented and discussed in section~\ref{sec:Results}. Section~\ref{sec:Summary} concludes this study. A companion study will put the results of this paper into context with the influence the waves exert on dynamic and thermodynamic conditions of the West African Monsoon.


\section{Methods}  \label{sec:Methods}

\subsection{Study area}
This study focuses on two \ang{5}-wide latitudinal bands in northern tropical Africa (Fig.~\ref{fig.map}): the Guinean (\ang{5}--\ang{10}N)  and Sahelian band (\ang{10}--\ang{15}N). For a more detailed analysis, two boxes in West Africa were defined where many rain gauges are available: the Guinea Coast (\ang{5}W--\ang{5}E, \ang{5}--\ang{10}N) and the West Sahel (\ang{5}W--\ang{5}E, \ang{10}--\ang{15}N). The analysis was stratified in two seasons: the full monsoon ranges from July to September; the three months before the onset of the monsoon (April to June) and October are collectively labeled as the transition season \citep{Sultan.2003b,Thorncroft.2011}; combining both, the extended monsoon season stretches from April to October.

\subsection{Data}

The modulating impact of tropical waves on rainfall over northern tropical Africa was examined in three different rainfall datasets. The Tropical Rainfall Measuring Mission (TRMM) 3B42 V.7 precipitation dataset is arguably the most accurate gridded rainfall product for the tropics \citep{Maggioni.2016}, with a record long enough for climatological  studies. TRMM 3B42 is a gauge-adjusted combined microwave-IR precipitation estimate \citep{Huffman.2007}. It has been used in several studies of tropical waves (e.g.,~\citealt{Yasunaga.2012,Lubis.2015}). The 3-hourly product has a spatial resolution of \ang{0.25}$\times$\ang{0.25}. The analyzed time period ranges from 1998 to 2016. With full spatial coverage in the tropics and a high temporal and spatial resolution, the TRMM dataset is a preferable dataset for the study of tropical waves in a rain gauge sparse environment such as tropical Africa. 

For a 33-year period from 1981 to 2013, the Climate Hazards Group InfraRed Precipitation with Station data V.2 (CHIRPS) provides daily, gauge-calibrated, infra-red based precipitation estimates \citep{Funk.2015}. The dataset has a resolution of \ang{0.25}$\times$\ang{0.25} and is available over land masses only. The daily CHIRPS product that was used is disaggregated from five-day accumulated values. Due to the disaggregation, CHIRPS can only be used with care at daily timescales and will likely underestimate the variance when compared to rain gauge observations (C.~Funk, 2017, personal communication). \blue{At \ang{0}E, the daily values range from 00 to 00UTC+1d.} This dataset has the advantage of a long record and a high spatial resolution. On the other hand, it has a comparatively low temporal resolution and is only available over the continent. 

\begin{figure*}[t]
	\centering
	\noindent\includegraphics[width = 0.75 \textwidth ]{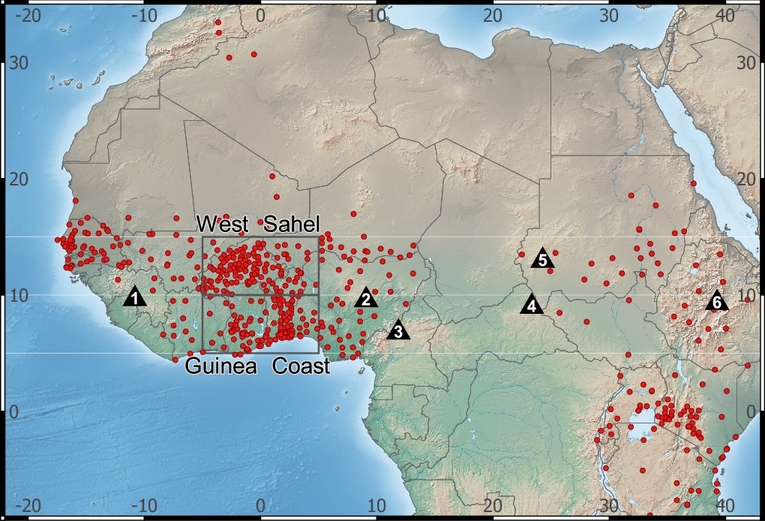}\\
	\caption{Topographic map of the study area. Available stations with a minimum of 50\% observations during 1981 to 2013 are indicated by a red dot. The two boxes show the location of the Guinea Coast and West Sahel used in this study. Black triangles show prominent orographic features that are discussed in the text: 1.~Guinea Highlands, 2.~Jos Plateau, 3.~\blue{Cameroon Line}, 4.~Bongo Massif, 5.~\blue{Darfur Mountains}, and 6.~Ethiopian Highlands.}\label{fig.map}
\end{figure*}

In situ rain gauge measurements are essential to validate the results obtained from satellite observations. The Karls\-ruhe African Surface Station Database (KASS-D) provides daily measurements from an extended rain gauge network. The 24-h accumulation period ranges for most stations from 06~UTC to 06~UTC. Stations were required to have at least 50~\% of observations during the period from 1981 to 2013 to match with the analyzed time period of the CHIRPS dataset. Additionally, stations with less than 1~\% of non-zero observations were excluded, resulting in a total number of 524 stations (Fig. \ref{fig.map}).

A good proxy for tropical convection is OLR (see review in \citealt{Arkin.1989}). OLR has been successfully applied in other studies of tropical waves (e.g.,~\citealt{Wheeler.1999,Roundy.2012}). Therefore, the daily interpolated National Oceanic and Atmospheric Administration (NOAA) OLR dataset \citep{Liebmann.1996} was used to filter the waves in the period from 1981 to 2013. The spatial resolution of this dataset is \ang{2.5}$\times$\ang{2.5}. To demonstrate the effect of orography, we used the \mbox{5-minute} gridded ETOPO5 elevation dataset \citep{NOAA.1988}.

\subsection{Wave filtering}
\begin{table*}
	\caption{Analyzed tropical waves with their corresponding wave characteristics. The filter settings for the period $\omega$, the wave number $k$, and the shallow water depth $h$ were used to extract the wave signal. Corresponding references are given.}
	\begin{center}
		{\small
		\begin{tabulary}{14cm}{ l L l l l l L}
			\toprule
			Acronym & Wave & Direction & $\omega$ (days) & $k$ & $h$ (m) & Source \\ 
			\midrule
			MJO & Madden-Julian Oscillation & Eastward & 30 -- 96 & 0 -- 9 & - & Roundy and Frank (2004) \\
			ER & Equatorial Rossby wave & Westward & 9 -- 72 & 1 -- 10 & 1 -- 90 & Kiladis et al. (2005, 2009) \\ 
			MRG & Mixed Rossby-gravity wave & Westward & 3 -- 8 & 1 -- 10 & 8 -- 90 & Wheeler and Kiladis (1999) \\
			Kelvin & Kelvin wave & Eastward & 2.5 -- 20 & 1 -- 14 & 8 -- 90 & Wheeler and Kiladis (1999) \\ 
			TD/AEW & Tropical disturbance/
			African Easterly Wave & Westward & 2.5 -- 5 & 6 -- 20 & - & Lubis and Jacobi (2015) \\ 
			EIG & Eastward inertio-gravity wave & Eastward & 1 -- 5 & 0 -- 14 & 12 -- 50 & Yasunaga and Mapes (2012) \\
			\bottomrule
		\end{tabulary} }
	\end{center}
	\label{tab.wave_characteristics}
\end{table*}
Six different wave types were analyzed in this study, sorted according to their scale: MJO, ER, MRG,  Kelvin, TD, and EIG (n=0, hereafter simply called "EIG") waves. WIG and higher \blue{meridional mode} EIG ($n>1$) could not analyzed \blue{because their frequency is too high to be analyzed using daily data}. The activities of the tropical waves were filtered in the wavenumber-frequency spectrum following the method developed by \citet{Wheeler.1999}. The filter settings for wavenumber, frequency, and equivalent depth ranges are obtained from previous studies on tropical waves studies (Table~\ref{tab.wave_characteristics}). \blue{No symmetric/antisymmetric decomposition has been done as the analyzed bands are entirely north of the equator. This results in only a small overlap of the filter bands of Kelvin and EIG waves, and of TD and MRG waves that is likely not important. It has to be noted that the wave signal, as captured by the filter bands, can be contaminated by random noise. The obtained signal must be interpreted as an overestimate of the wave modulation since wave motions generally stay out against the background noise by less than \SI{30}{\percent} \citep{Wheeler.1999}.}

\blue{The modulation of precipitation by tropical waves was investigated for two periods: (a) for the shorter period 1998 to 2016, the influence of tropical waves was analyzed in the TRMM dataset; (b) for the longer period 1981 to 2013, CHIRPS and KASS-D were examined. Table~\ref{tab.datasets} shows the structure of the results section and lays out which dataset has been used in which subsection. For the analysis of the rainfall patterns and modulation intensities of each wave, OLR is used by default for the wave filtering to derive the local wave phase. Precipitation anomalies or quantile deviations are then projected onto this composite. TRMM data have also been used in place of OLR to determine the wave phase (supplemental material) to assess the robustness of the composite. For the analysis of the contribution of each wave to the rainfall variability and the wave interaction, TRMM has been wave-filtered directly because of its global coverage, which is required for the wave-filtering.

The filtering was done using the NCAR Command Language (NCL) \textit{kf\textunderscore filter} function. This function is based on a Fast Fourier Transform (FFT), which does not allow missing values in the entire analyzed latitudinal band. For the longer period from 1981 to 2013, we rely on the interpolated OLR dataset as a proxy for precipitation, although, the relationship between OLR and precipitation is weak over West Africa (see Fig. S1). CHIRPS is available only over the continents and could thus not be used for filtering. The 3-hourly TRMM product exhibits minor observational gaps.} Over Africa, these gaps are negligible with an average amount of missing data of about \SI{0.005}{\percent} to \SI{0.05}{\percent}. All missing values in TRMM were filled with zeros. The effect of this primitive gap filling method for the wave filtering is expected to be negligible over Africa where gaps are rare. As a side remark, more caution should be applied when the 3-hourly TRMM product is used for wave filtering over the Maritime Continent for the period prior to 2007, where about 1\% of the data is missing due to downtimes of a geostationary IR-satellite. 

\begin{table*}
	\caption{Structure of the study and use of datasets. The table lists the subsections in the result part and outlines which datasets have been used to filter for the tropical waves and which precipitation datasets have been subsequently plotted in the corresponding figures.}
	\begin{center}
		{\small
		\begin{tabulary}{14cm}{  l  L  L  L  l  }
			\toprule
			Section & Topic & 1. Wave filtered dataset & 2. Plotted dataset (raw, composite, correlation) & Figs. \\ 
			\midrule
			3.a & Mean climate & - & TRMM & 3 \\ 
			3.b & Total variance of wave activity & TRMM & TRMM & 4 \\ 
			3.c & Modulation patterns & OLR & CHIRPS, KASSD & 5-6 \\ 
			3.d & Modulation intensity & OLR & CHIRPS & 7 \\ 
			&  & OLR & KASSD & 7-8 \\ 
			&  & TRMM & TRMM & 7 \\ 
			3.e & Relative contribution to precipitation & TRMM & TRMM & 9-11 \\ 
			3.f & The role of orography & TRMM & TRMM & 9-11 \\ 
			3.g & Wave interactions & TRMM & TRMM & 12 \\ 
			\bottomrule 
		\end{tabulary} }
	\end{center}
	\label{tab.datasets}
\end{table*}

The filtering method has no constraints for the associated circulation. Wave signals can be contaminated by physically different phenomena, missed by the filter due to Doppler shifting, or double counted due to the partial overlap of wavenumber frequency spectra. Other filters have been developed that are based on the horizontal structure functions obtained from equatorial wave theory \citep{Yang.2003}, use wavelet analysis \citep[e.g.][]{Kikuchi.2010,Roundy.2018} or 3D normal mode functions \citep{Castanheira.2015}. Nonetheless, \blue{the here applied method has been successfully used in previous studies over Africa (e.g. \citealt{Mounier.2007,Mounier.2008,Janicot.2009,Janicot.2010,Ventrice.2013,Mekonnen.2016}). In a subsequent article it will be shown that} composites of dynamical fields using \blue{this method} are able to reproduce the circulation patterns that are expected from wave theory.

\blue{As a metric for the MJO activity, several EOF-based global indices have been proposed, such as the Real-time Multivariate MJO (RMM, \citealt{Wheeler.2004}) index and the OLR-based MJO Index (OMI, \citealt{Kiladis.2014}). These indices have the advantage of measuring the global propagation of the MJO and being less affected by background noise. Over Africa conversely, the MJO manifests itself rather as a standing signal \citep{Alaka.2012,Alaka.2014,Alaka.2017}. A local wave filtering, as applied in this study, will thus mainly depict the standing signal over Africa and might miss the global propagative signal as seen over the Indian and Pacific Oceans.} 

\begin{figure*}[t]
	\centering
	\noindent\includegraphics[width = 0.95 \textwidth ]{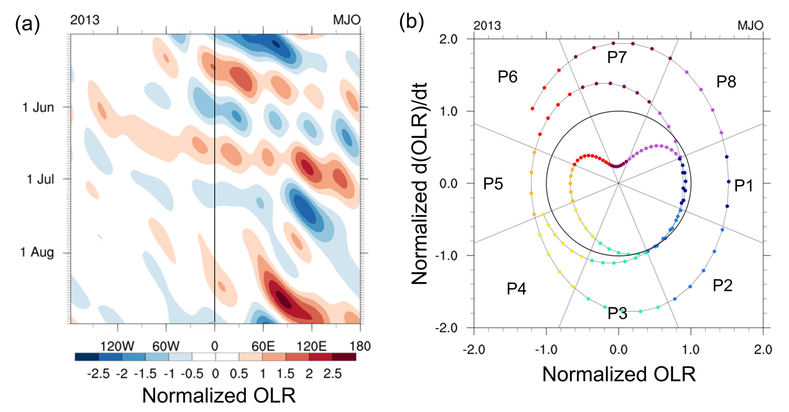}\\
	\caption{Construction of composite plots. (a) Example of a Hovmoeller diagram of an MJO filtered normalized OLR signal. Negative anomalies correspond to wet conditions. (b) Example for the MJO in a phase diagram. Plotting the wave filtered signal at \ang{0}E (black line in (a)) against the time derivative at \ang{0}E, results in a phase diagram of the local wave activity. The phase space is divided into eight phases (coloring); dates when the amplitude is less than one standard deviation are considered to be inactive. }\label{fig.phase}
\end{figure*}

\subsection{Composite analysis}
\blue{Composites of the waves were constructed to investigate the propagation and structure of modulation patterns by averaging rainfall anomalies over all dates during the extended monsoon season when the wave is in a specific phase based on the OLR signal (Fig.~\ref{fig.mapER}--\ref{fig.mapAll}). A local wave phase and amplitude is defined following the method initially developed by \citet{Riley.2011} for the MJO. The filtered wave signal and its corresponding time-derivative are calculated for a reference location in West Africa at \ang{0}E and \ang{5}--\ang{15}N (Fig.~\ref{fig.phase}a). Both fields are standardized by dividing by the standard deviation of the respective field. The phase diagram is then split in eight equal phases (Fig.~\ref{fig.phase}b). Phase 1 refers to a local dry phase over the reference location, phase 5 to the wet phase; during phase 3 and 7 the wave is in its neutral phase, while the rest of the phases are transition phases. During dates when the amplitude of the wave activity is less than one standard deviation, the wave is considered to be inactive. For a more detailed description of the calculation of wave phases, the reader is referred to \citet{Yasunaga.2012} and \citet{vanderLinden.2016}, where this approach is also applied to study the local influence of tropical waves. For all waves about 400--450 days are available in each phase. \blue{Additional maps showing the anomalies based on TRMM filtered waves instead of OLR are available in the supplementary material (Figs.~S7--\blue{S12}). These maps are not as smooth as the plots obtained from CHIRPS, because the sample size is smaller due to the shorter period from 1998--2013; they have, however, the advantages that the wave signal is directly filtered from precipitation instead of OLR and that they include rainfall anomalies over the ocean. As a side remark, it was tested whether OLR-filtered TRMM composites are less noisy. However, the noisiness was not removed and the general patterns did not change significantly (not shown).} Statistical significance of the anomalies in the composite analysis (Figs.~\ref{fig.mapER}--\ref{fig.mapAll}) is calculated using non-parametric bootstrapping. Significance was tested at a \SI{5}{\percent} level using a sample of 1000 repetitions. \blue{Using a compositing method, the background noise will be mostly removed and the resultant patterns will reflect at least to first order the linear relationship to the analyzed wave.}

\subsection{Normalization of precipitation}
\blue{The composite maps of rainfall require a prior normalization.} Northern tropical and subtropical Africa features a wide range of different climates. Annual precipitation ranges from more than 2500 mm yr$^{-1}$ in coastal perhumid areas of western and central Africa to less than 100 mm yr$^{-1}$ in the hyperarid areas in the Saharan desert \citep{Fink.2017}. In order to make the effect of tropical waves on precipitation anomaly patterns comparable for these very diverse climates, precipitation values need to be normalized. \blue{Using total rainfall anomalies, the modulation in drier regions would not be visible.} Indices like the Standardized Precipitation Index (SPI) \citep{McKee.1993} cannot be calculated on the daily timescale. Here, we propose to use a quantile based approach instead, in order to derive normalized daily precipitation anomalies against a climatological reference. An adjusted definition of quantiles was used to account correctly identical values and, in particular, the many zero observations in hyperarid regions. We define the quantile $q$ for an observation $y$ and set of climatological observations $O$ of size $N$ as
\blue{\begin{align*}
		q(y):=\frac{|\lbrace a \in O | a < y \rbrace | + \frac{1}{2} | \lbrace b \in O | b = y \rbrace | }{N}.
\end{align*}}

\blue{In other words, $q$ gives the percentage of observations $a$ that are smaller than $y$ plus half the percentage of observations $b$ that are equal to $y$, both relative to the total number of observations. The later part is done to equally weigh observations with the same value. By definition, the  median of the climatology equals to $\tilde q=50 \%$.} Anomalies are \blue{then} calculated as deviations from the median, resulting in quantile anomalies as measured in percentiles $a = (100q - 50) \%$. This approach allows to compare precipitation anomalies of different climates.

To derive a reasonable sample for the reference climatology, the concept of an extended probabilistic climatology is used (see \citealt{Vogel.2018} for a more detailed discussion). To increase the sample size and reduce the sampling error, the climatology was defined as a set of observations $\pm$ 7 days around the respective day of the year. A sample of $\pm$ 7 days reasonably minimizes the sampling error and bias (\citealt{Vogel.2018}, Fig.~S1 in their supplementary material). Larger windows increase the bias due to different climatological settings at the beginning and end of the window. 

\subsection{\blue{Modulation intensity}}
\label{subsec:intens}
\blue{Daily rainfall anomalies were computed for the period from 1981 to 2013 for CHIRPS and KASS-D and for the period from 1998 to 2013 for TRMM. Then, the average anomaly during each phase was plotted. }For \blue{the two focus} regions in West Africa, \blue{the modulation intensity was measured for all three precipitation datasets}. 112 stations are located in the Guinea Coast, whereas the West Sahel box includes 129 stations (Fig.~\ref{fig.map}). The wave activity in both regions was calculated from OLR for the Guinean (\ang{5}--\ang{10}N) and Sahelian (\ang{10}--\ang{15}N) zones, respectively, at \ang{0}E. \blue{Most of the stations measure from 06 to 06 UTC+1d. In order to remove the time lag of six hours, we shifted the curve for KASS-D by the estimated length of six hours measured in wave phases: \[\Delta P = \frac{\SI{6}{\hour}}{T \text{[in h]}/8},\]  where $T$ equals to the mean period of the filtered wave signal as determined by the MATLAB function \textit{meanfreq()}.}

\blue{The distribution of modulation intensity as seen by different stations was also calculated for both seasons.} The mean modulation intensity for each station was calculated by subtracting the mean precipitation in the wettest and driest phase of the eight phases defined above. Due to potential shift between OLR and the observed station precipitation, the wettest phase is allowed to fall into phases 4--6, whereas the driest phase can lie in phases 8--2.

Kernel density estimation (KDE) has been used to visualize how the modulation intensities vary for the different stations. \blue{Commonly, histograms are used to visualize discrete sample of observations. Yet, histograms have disadvantages when visualizing probability density functions (PDFs) of finite samples: They are discontinuous and their shape highly depends on the choice of bin width; additionally, the interpretation of several overlying histograms is rather difficult. A non-parametric alternative to histograms is KDE. In order to create a smooth and, continuous PDF, each observation is weighted with a kernel function.} As kernel function we chose a normal distribution. Results did not differ significantly with a different choice of kernel. The bandwidth $h$ of the kernel was calculated following the rule-of-thumb by \citet{Freedman.1981}:
\[h= \frac{2 \cdot IQR}{N^\frac{1}{3}} ,\]
where $IQR$ is the inter-quartile range and $N$ the sample size. An illustrative review explaining KDE and describing the advantages of this method can be found in \citet{GonzalesFuentes.2015}.

\subsection{\blue{Correlation analysis}}
\blue{The influence of the different tropical waves is expected to depend on the timescale. For example, daily rainfall variability will be mainly driven by waves with a high frequency such as TDs, whereas weekly accumulated rainfall will be mainly determined by low-frequency waves such as the MJO.} In order to test how much of the rainfall variability on different timescales can be explained by the modulation of tropical waves the linear correlation coefficient between the rainfall variability on the specific timescale and the wave filtered signal is calculated.  In a first step, the variability on different timescales was calculated, applying a running mean of one, three, seven, and 20 days as well as \ang{1} in longitude \blue{on the mean TRMM precipitation in the Guinean and Sahelian bands for the transition and full monsoon seasons from 1998 to 2016. This way, the variability on the respective shorter timescales is removed and only the variability on longer timescales is retained. The data are then correlated at each longitude with the wave filtered TRMM signal in the same band. The squared correlation coefficients estimate the explained variance of the rainfall by each wave in this band and during the respective season.}

\blue{In order to visualize how the correlation varies within the Guinean and Sahelian bands, the correlations of all waves at each longitude were stacked on top of each other for all timescales (Figs.~\ref{fig.rhoGuineaTransition}--\ref{fig.rhoSahelFull}). The sum of all correlations should be considered as a maximum percentage that can be explained by tropical waves, as some of the wave bands have small overlaps, and no wave interactions are taken into account.  Random noise does not systematically affect the linear correlation, so the correlation shows to first order how much the wave modes contribute to total rainfall variability.}
	
\blue{Statistical significance of the correlation coefficients was calculated using a one-sided t-test at a significance level of $p<0.05$.} As both time series are autocorrelated in time, a reduced number of degrees of freedom needs to be used to test the significance. For two Gaussian distributed autocorrelated timeseries $x$ and $y$ of a length $N$, the reduced degree of freedom $DF_{eff}$ is calculated using

\[DF_{eff}=\frac{N}{1+2\sum_{l=0}^{N-1} \frac{N-l}{N} K^\ast_x(l) K^\ast_y(l)}+1,\]

where $K^\ast_x(l)$ and $K^\ast_y(l)$ are the autocorrelation functions of both time series (Taubenheim, 1974 cited by \citealt{Fink.1997}).

\subsection{Wave interactions}
\label{subsec:interaction}
The interactions between different types of tropical waves were tested. The mean filtered TRMM precipitation  was calculated for the Guinea Coast (\ang{5}--\ang{10}N, \ang{0}E) and the West Sahel (\ang{10}--\ang{15}N, \ang{0}E) for the extended monsoon season. Dates when the filtered precipitation exceeded $ \pm 1$ standard deviation were defined as wet and dry phases of the wave. On average, about 4000 observations fall in each phase for all waves. The average wave-filtered signal in the wet and dry phases is called the primary modulation.

Next, it was analyzed how much a second wave modifies the primary modulation. Therefore, the mean wave-filtered signal of the primary wave was calculated under the condition that a second wave was either in a positive or negative phase. For each possible wave superposition, the sample size was about 500--800 cases. The difference of the mean wave signal during the passage of the second wave to the mean modulation by the primary wave is called the secondary modulation. It should be noted that these values are to be interpreted as the activity in the specific wave spectrum rather than as physical rainfall anomalies. Significant differences were tested using a t-test for samples of different variances with a significance level of \SI{5}{\percent}.


\section{Results and discussion}  \label{sec:Results}
 
\subsection{Mean climate}
	
	\begin{figure*}[t]
		\centering
		\noindent\includegraphics[width = 0.75 \textwidth]{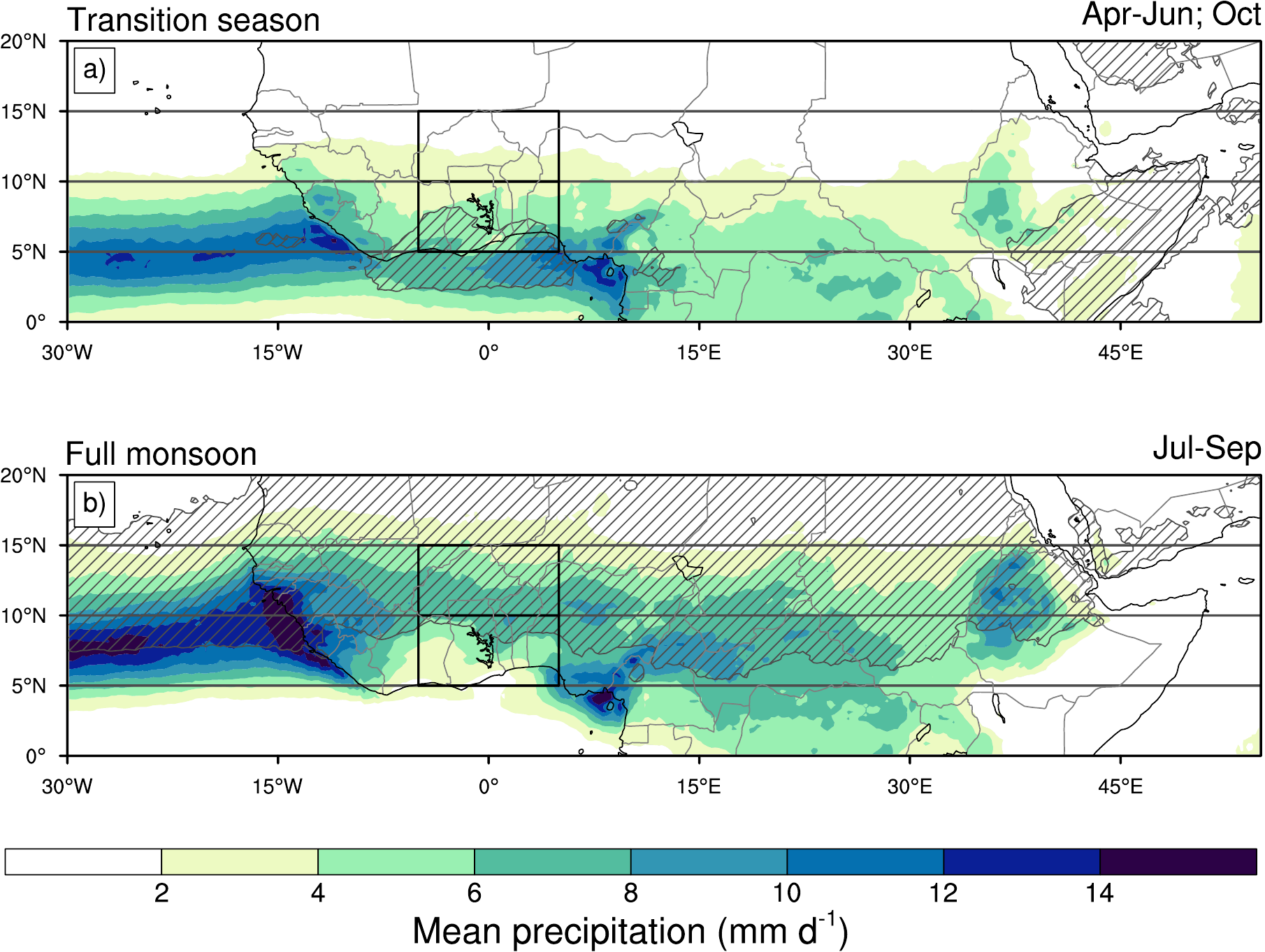}\\
		\caption{Mean seasonal precipitation from TRMM observations (1998-2016) during (a) transition season (April - June; October) and (b) the full monsoon (July - September). Areas with more than 50 \% of annual rainfall during the respective season are hatched. Black lines denote the boxes and bands as used in subsequent analyses.}		
		\label{fig.trmm}
	\end{figure*}
	
	The African monsoon system dominates the circulation and distribution of rainfall over northern tropical Africa. During the transition season (Apr.--Jun.; Oct.), when the monsoon has not yet fully started or already ended, rainfall is concentrated over the coastal regions around \ang{5}N (Fig.~\ref{fig.trmm}a). \SI{49}{\percent} of the annual precipitation or a total of ca.~\SI{600}{\mm} falls in the Guinea Coast box during this season as outlined in Fig.~\ref{fig.trmm}. In contrast, the West Sahel box receives \SI{28}{\percent} of the annual precipitation or \SI{230}{\mm} in the same season. With the start of the African monsoon, the main rainfall band moves northwards to about \ang{10}N during the full monsoon season from July to September (Fig.~\ref{fig.trmm}b). Then, \SI{71}{\percent} of the annual precipitation or \SI{530}{\mm} falls in the West Sahel; the coastal regions of West Africa experience the little dry season. Only \SI{37}{\percent} of the annual precipiation or \SI{450}{\mm} falls during this season in the Guinea Coast.  Due to the West African Monsoon, the northward shift is stronger over the land than over the ocean.	Rainfall over elevated regions is generally higher than in the lowlands. Prominent orographic include the Guinea Highlands, the Jos Plateau, the \blue{Cameroon Line}, the Bongo Massif, the \blue{Darfur Mountains}, and the Ethiopian Highlands as shown in Fig.~\ref{fig.map}. It will be demonstrated in subsection~\ref{sec:Results}.\ref{subsec:relCon2PcP} that these orographic features also play a role for how precipitation is modulated by tropical waves.

\subsection{Total variance of wave activity}
	\begin{figure*}
	\centering
	\noindent\includegraphics[height = 0.85 \textheight]{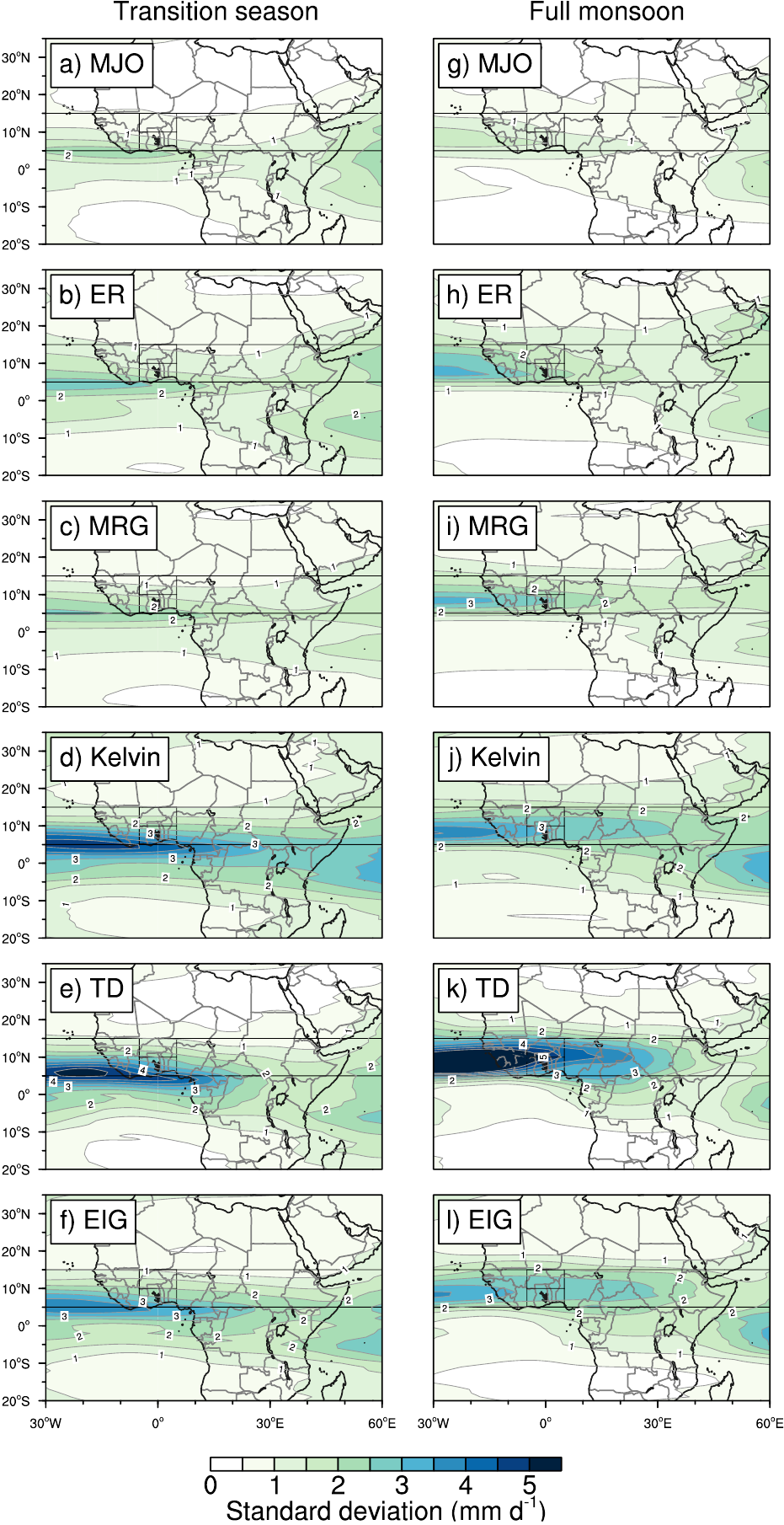}\\
	\caption{Standard deviation of TRMM precipitation (1998--2016) within the specific wave domains for the transition season (a-f) and full monsoon (g-l). Black lines denote the boxes and bands as used in the analysis.  For abbreviations of wave names, see Table~\ref{tab.wave_characteristics}.}\label{fig.trmmwavevar}
	\end{figure*}
	In a first step, it was analyzed where and how much rainfall variability falls into the specific wave spectra during the transition and full monsoon seasons. For this purpose, Fig.~\ref{fig.trmmwavevar} shows the standard deviation of wave filtered TRMM precipitation signals during the period from 1998 to 2016. The magnitude of rainfall variability differs in the analyzed wave spectra. \blue{CCEW do not show highest variability at the equator as might be expected from the shallow-water equations}. Rather, the highest contribution of tropical waves to rainfall variability is directly linked to the mean precipitation patterns and lies where the seasonal rainfall maximum is located. \blue{It has to be noted that the background noise significantly contributes to the variance within the different bands and therefore bands with more variance in the raw red spectrum naturally exhibit higher variance in the unfiltered data as well (cf. Figs. 1 and 2 in \citealt{Wheeler.1999}).} The highest variability can be observed for TD and Kelvin waves. During the transition season, TD and Kelvin waves have a comparable variability with a standard deviation of about 3--\SI{4}{\mm\per\day} at the Guinea Coast. MJO, ER, and MRG are of similar variability with a maximum standard deviation of about \SI{2}{\mm\per\day}. The variability in the EIG band is slightly stronger with a maximum standard deviation of about \SI{3}{\mm\per\day}. 
	 
	During the full monsoon in contrast, the TD signal is by far the most dominant source of rainfall variability. The intensity in the TD band increases from the Ethiopian highlands towards maximum standard deviation of more than \SI{6}{\mm\per\day} over the west coast of Africa. The variability for the other waves is comparable with a maximum standard deviation of about 2--\SI{3}{\mm\per\day}. During this season, ER and MRG waves exhibit slightly more variability, whereas the MJO and Kelvin waves are slightly less active.

	Several authors have compared the different waves on a global scale \blue{\citep{Wheeler.1999,Roundy.2004,Kiladis.2009,Huang.2011,Schreck.2013,Lubis.2015}}. Some of these only provide results for the entire year \blue{\citep{Wheeler.1999,Kiladis.2009}}. The global maps for the summer and spring season in \citet{Roundy.2004} using OLR roughly agree with the results in Fig.~\ref{fig.trmmwavevar}. The drawback of the studies that filter the waves from OLR or brightness temperature as indicators for deep convection is the low correlation of OLR with daily precipitation over Africa, which is generally less than 0.3 (Fig.~S1, supplementary material). \citet{Huang.2011}, and \citet{Lubis.2015} show the seasonal cycle of different waves demonstrating that the maximum of TD and MRG-activity is during the boreal summer, ER peak during fall and winter, whereas Kelvin waves show the highest activity during boreal spring. \citet{Lubis.2015} filter the waves from TRMM precipitation. Their results of the activity of different waves during the extended monsoon season are comparable with the presented results. \citet{Skinner.2013} compare in their Fig. 8 the variance of precipitation in different tropical wave spectra over West Africa during June to October in different general circulation models and observations from the Global Precipitation Climatology Project (GPCP). The overall picture is similar: the wave type explaining most variance is TD followed by ER and Kelvin waves. It should be noted that there is considerable inconsistency between the different models and observations, stressing the need for improved representation of the waves in NWP models.
	
\subsection{Modulation patterns}

\begin{figure*}
	\centering
	\noindent\includegraphics[height = 0.75 \textheight ]{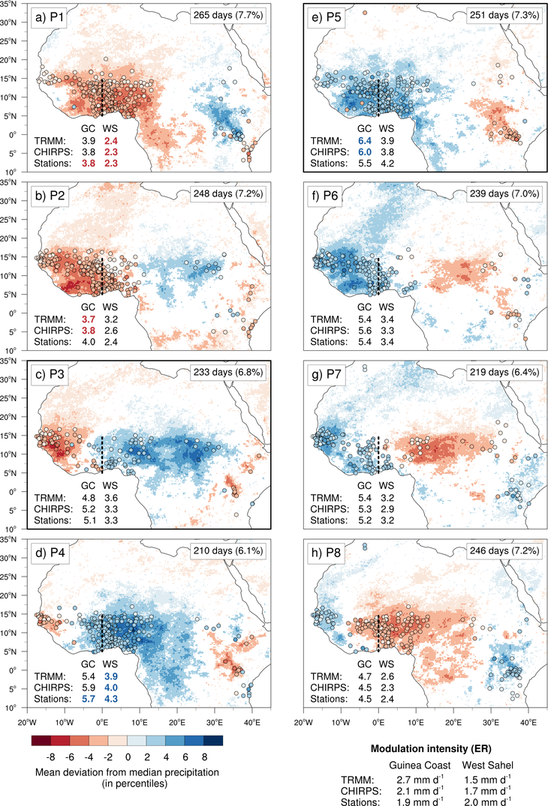}\\
	\caption{Rainfall composite for days with significant ER wave signal\blue{, based on OLR anomalies} over \ang{5}-\ang{15}N, \ang{0}E (dashed line), for CHIRPS (\blue{shading}) and KASS-D (circles, both 1981--2013) \blue{during the extended monsoon season (April - October)}. As calculated with a bootstrap test, non-significant anomalies ($p>0.05$) are left white for the CHIRPS data, KASS-D stations that are non-significant are not shown. The number of days used for the composites and the fraction of days per season during each phase is given in the upper right of the plot. The numbers in the lower left part of the subpanels states the mean observed rainfall in all three rainfall datasets for the given phases (TRMM period: 1998--2013) within the Guinea Coast and West Sahel (black boxes). The modulation intensity, measured as the difference between the mean rainfall amount in the wettest (blue number) and driest phase (red number), is summarized at the right bottom.}\label{fig.mapER}
\end{figure*}

	The discussion of Fig.~\ref{fig.trmmwavevar} already indicated that different wave types modulate rainfall amounts to a different degree. In a next step, the effect of tropical waves on precipitation patterns and modulation intensity during the extended monsoon is analyzed. The modulation patterns of the ER wave are shown for all eight phases and will be discussed in detail because of the notable influence far into the subtropics (Fig.~\ref{fig.mapER}). For the remaining waves only the wet phase (P5) and one neutral phase (P3) will be shown; the full figures showing all phases \blue{and maps with precipitation anomalies as measured and filtered by TRMM} can be found in the supplementary material (Figs.~S2--S12).
	
	Although the composites are based on a local wave filtering for a reference location at \ang{0}E, \ang{5}-\ang{15}N, the ER wave shows significant and spatially coherent modulation patterns over most of northern tropical Africa. The eight phases give a clear picture of the propagation of the ER wave. As an largely independent dataset, rain gauges measurements confirm the significant influence over the entire continent. The wave exhibits significant modulation patterns that reach up to the Mediterranean Sea and to East Africa. The approximate wave length is \SI{8000}{\km}; thus, it is dry over Central/East Africa when wet anomalies persists over West Africa and vice versa. \blue{The modulation patterns as measured by TRMM are very similar to CHIRPS but more noisy and not as pronounced (Fig.~S7); as expected from theory, weak symmetric signals at \ang{10}S can be seen in several phases over the Atlantic Ocean.} The modulation patterns of ER waves in \blue{\citet{Janicot.2010, Janicot.2011} and \citet{Thiawa.2017}} are remarkably similar to the results of this composite study. The slightly eastwards tilted precipitation pattern resembles tropical plumes regularly observed during the dry winter season \citep{Knippertz.2005,Frohlich.2013}. An analysis of the dynamical fields is needed to verify whether such a relation between tropical plumes and ER waves does in fact exist.
		
	The remaining waves also show modulation patterns influencing entire northern tropical Africa (Figs.~\ref{fig.mapAll}). The MJO modulates the precipitation over the entire African continent far into the subtropics up to the Mediterranean Sea and to East Africa (\blue{Figs.~\ref{fig.mapAll}a, f}, S2, \blue{and S8}). Over the west coast, South Sudan, and Central Africa, the modulation is weak. Over the complex East African terrain, the MJO seems to be out of phase. For the transition phases, a weak zonal dipole exists, with wet (dry) conditions over the tropical band and dry (wet) conditions over the Sahara. \blue{Weak anticorrelated wave patterns south of \ang{5}N can be seen over the Atlantic Ocean (Fig.~S8).} The tilted precipitation anomalies over the Sahara as seen for the ER, can also be observed for MJO waves, although this pattern is not as pronounced (P2 and P4, see Fig.~S2). \blue{Over the Atlas Mountains, precipitation is enhanced during the wet phases and reduced during the dry phases (cf. also Fig.~S8). The TRMM-filtered composites suggest that the signal over the Gulf of Guinea precedes the signal over the continent.}
	
	\begin{figure*}
		\centering
		\noindent\includegraphics[height = 0.85 \textheight ]{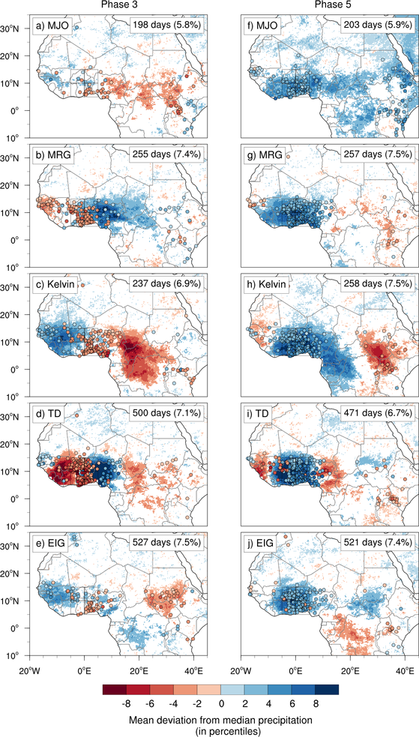}\\
		\caption{Same as Fig.~\ref{fig.mapER}, but for all waves and phase 3 (neutral phase, a-e) and phase 5 (wet phase, f-j). Plots showing all phases are provided in the supplementary material (Figs.~S2--S6). For abbreviations of wave names, see Table~\ref{tab.wave_characteristics}.}\label{fig.mapAll}
	\end{figure*}

	MRG waves are rather confined to the filtered band between \ang{5}-\ang{15}N and lose their clear pattern over East Africa (\blue{Figs.~\ref{fig.mapAll}b, g}, S3, \blue{and S9}). In the subtropics, only weak significant signals can be observed over the Maghreb region (P7, see Fig.~S3). The approximate wave length is $>$ \SI{9000}{\km}. An antisymmetric pattern in the Southern Hemisphere, as expected from theory, is \blue{observable with TRMM (Fig.~S9) but} not with CHIRPS due to the lack of data over oceans. Over Central and East Africa, the antisymmetric pattern is weak.
	
	Kelvin waves are an equatorial phenomenon. This is the reason, why the modulation also extends to \ang{10}S but not to the north of the filtered band at \ang{15}N (\blue{Figs.~\ref{fig.mapAll}c, h}, S4, \blue{and S10}). The precipitation patterns are well-defined and blur out over the eastern part of the continent. As a side remark, Kelvin waves might also locally modulate the land-sea breeze as indicated by a modulation which is out of phase at several coastal stations (P4 and P8, Fig.~S4).
	
	TD are most dominant over West Africa (\blue{Figs.~\ref{fig.mapAll}d, i}, S5, \blue{and S11}). The wave signal weakens east of \ang{10}E. The influence is also mostly confined to the filtered area between \ang{5}--\ang{15}N. A weak anticorrelation with precipitation in the Sahara can be observed. \blue{Over the Atlantic the precipitation pattern tilts in a northeast-southwest direction (Fig.~S11)}. The average wave length is about \SI{2500}{\km}.
		
	EIG waves show a weaker and more complex modulation pattern compared with the previous waves (\blue{Figs.~\ref{fig.mapAll}e, j}, S6, \blue{and S12}). The modulation is strongest over West Africa, equatorial Central Africa and South Sudan. Consistent with theory, antisymmetric rainfall patterns are evident in the Southern Hemisphere, although the anticorrelated rainfall regions are shifted northwards and centered around \ang{5}S. Very weak, significant anticorrelated precipitation anomalies further in the north could be an indication of a contamination by higher order (e.g., n=1, n=2) EIG waves. Further investigations are needed to better understand the involved mechanisms of triggering, propagation, and coupling of EIG waves to precipitation. Rain gauge observations do not match the CHIRPS dataset well due to a phase shift between both datasets, which will be the topic of the following section.
	
	\blue{A detailed comparison of modulation patterns and the associated circulation of all wave types with theory will follow in the second part of this study.}
	
\subsection{Modulation intensity}

\begin{figure}[p]
	\centering
	\noindent\includegraphics[width = 6.7cm]{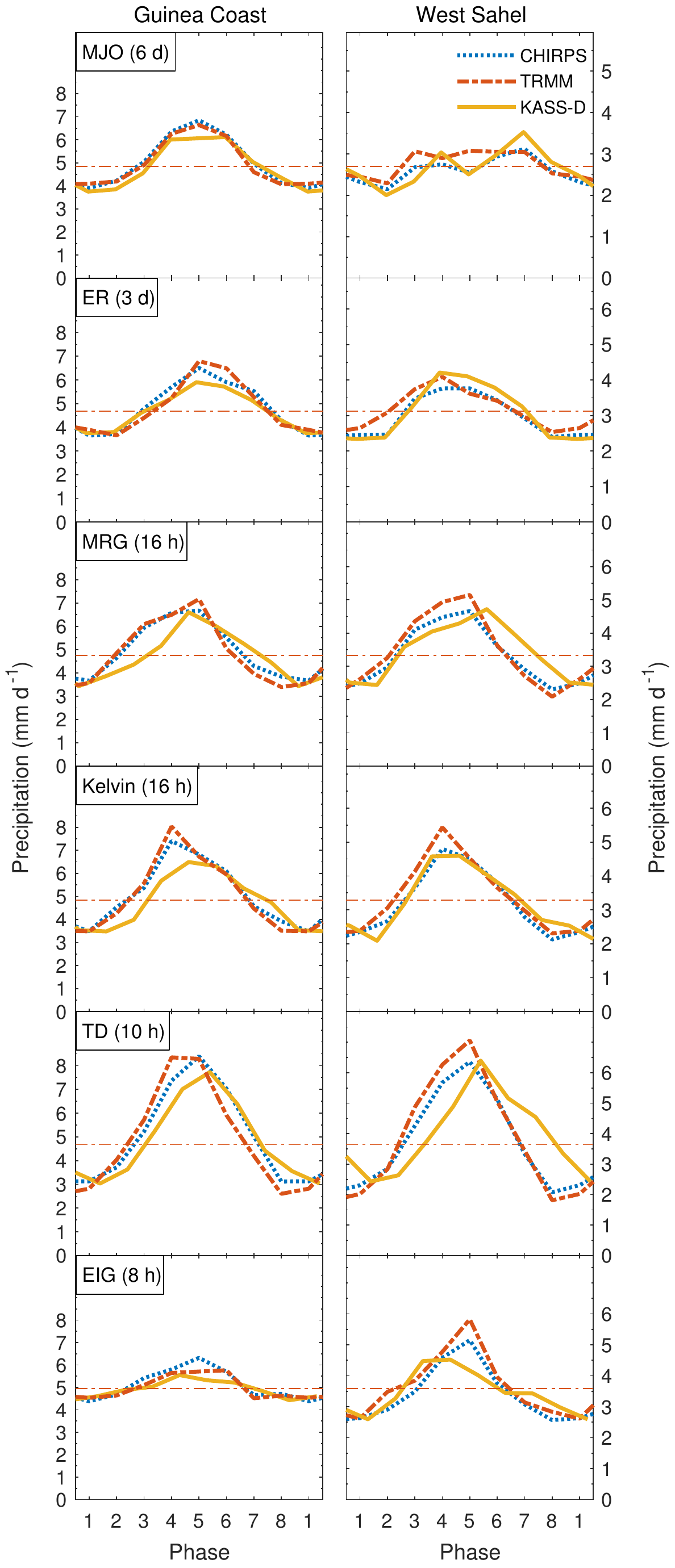}\\
	\caption{Mean precipitation in the eight phases in the Guinea Coast box (left side), and West Sahel box (right side) in CHIRPS (1981--2013, \blue{blue dotted line}), TRMM (1998--2013, red dashed line), and KASS-D (1981-2013, \blue{yellow solid line}) \blue{during the extended monsoon season (April - October). Because KASS-D lags the other datasets by \SI{6}{\hour}, the curve has been shifted using the mean period of each wave (in brackets). See section \ref{subsec:intens} for more detail}. Note that the y-axis has been scaled with the mean precipitation during all phases in the TRMM dataset, which is indicated by the horizontal red line.}
	\label{fig.PhasePrecip}
\end{figure} 
	
	\blue{The modulation intensity was measured in absolute rainfall. Figure~\ref{fig.PhasePrecip} shows the mean precipitation in the Guinea Coast and West Sahel boxes during the eight phases as measured by CHIRPS, TRMM, and KASS-D. Rainfall varies considerably between the different phases of all waves. Additionally, it can be seen that TDs and Kelvin waves have the strongest impact on absolute rainfall anomalies, followed by MRG and ER waves. Despite the known deficiencies of the datasets \citep{Funk.2015,Maggioni.2016}, the observed amplitudes agree well for all three datasets.}
	
	\begin{figure*}[p]
		\centering
		\noindent\includegraphics[width = 0.95 \textwidth ]{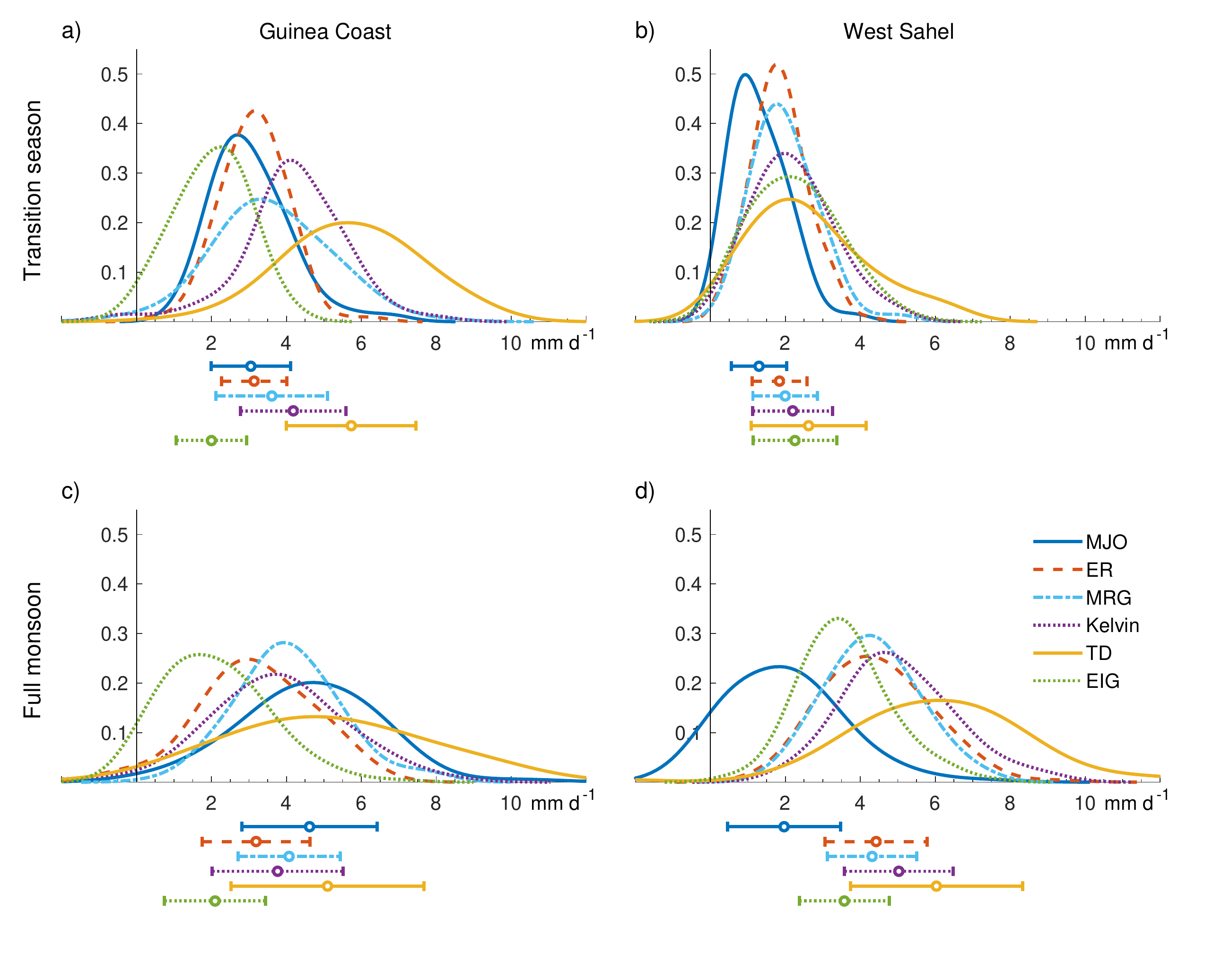}\\
		\caption{Modulation of rainfall by tropical waves in the Guinea Coast (\ang{5}W-\ang{5}E, \ang{5}-\ang{10}N) and the West Sahel (\ang{5}W-\ang{5}E, \ang{10}-\ang{15}N) during the transition season (Apr.--Jun.; Oct., top) and the full monsoon season (Jul.--Sep., bottom) \blue{as measured by rain gauges in KASS-D}. Dates in wet and dry phases of the tropical waves were obtained by filtering OLR in a Guinean (\ang{5}-\ang{10}N) and Sahelian band (\ang{10}-\ang{15}N) from 1979-2013. As a metric for the modulation intensity, the difference of the mean precipitation during wet phases and dry phases was calculated for all stations (Guinea n=106, Sahel n=117). From these stations an empirical probability density function was derived using kernel density estimation. Lines below the axis show  mean and standard deviation of the corresponding modulation intensity for each wave. For abbreviations of wave names, see Table~\ref{tab.wave_characteristics}.}
		\label{fig.StationsPDF}
	\end{figure*}

	In a more detailed analysis, gauge observations will be analyzed for the transition season and the full monsoon season. Figure~\ref{fig.StationsPDF} shows the distribution of mean modulation for all stations in the West Sahel and Guinea Coast box as the difference of modulation intensities during wet and dry phases of the waves. \blue{As a side-remark, it can be noted that negative modulation intensities can be observed, which means that less rain at single stations falls during the wet phase and vice versa. These cases are likely related local factors such as orography (cf.  Figs.~\ref{fig.mapER}--\ref{fig.mapAll} and section \ref{sec:Results}.\ref{subsec:orogr}) and random errors.} The corresponding plot in quantile anomalies can be found in the supplementary material (Fig.~S13).
	
	During the transition season, rainfall at the Guinea Coast is strongly modulated by tropical waves (Fig.~\ref{fig.StationsPDF}a). Descending in order, the mean modulation intensities are 5.7~(TD), 4.2~(Kelvin), 3.6~(MRG), 3.1~(ER), 3.0~(MJO), and \SI{2.0}{\mm\per\day}~(EIG). The standard deviations lie between \SI{0.8}{\mm\per\day} for ER and \SI{1.7}{\mm\per\day} for TD. The West Sahel is still relatively dry during the transition season (Fig.~\ref{fig.trmm}a). Therefore, the influence of tropical waves is weak, amounting to only 1.3--\SI{2.6}{\mm\per\day} for all waves (Fig.~\ref{fig.StationsPDF}b).
	
	In the full monsoon season, the modulation at the Guinea Coast is similar to the transition season (Fig.~\ref{fig.StationsPDF}c). The mean modulation intensities amount to 5.9 (TD), 4.6 (MJO), 4.1 (MRG), 3.7 (Kelvin), 3.2 (ER), and \SI{2.1}{\mm\per\day} (EIG). A noteworthy difference to the transition season is a stronger modulation by the MJO, although the MJO activity itself is weaker during the full monsoon (Fig.~\ref{fig.trmmwavevar}). The standard deviations are higher than during the transition season due to the high gradient of rainfall at the Guinea Coast during the monsoon season (Fig.~\ref{fig.trmm}). Surprisingly, all CCEWs are as strong or even stronger in the West Sahel (Fig.~\ref{fig.StationsPDF}d) than over the Guinea Coast (Fig.~\ref{fig.StationsPDF}c), although the strongest influence of equatorial waves would be expected near the equator from theory. The mean modulation intensities are 6.0 (TD), 5.0 (Kelvin), 4.4 (ER), 4.3 (MRG), 3.5 (EIG), and \SI{2.0}{\mm\per\day} (MJO). For both regions, the largest standard deviation is seen in the modulation intensity for TD. Several stations record a modulation intensity of more than 8 mm/d. 

\subsection{Relative contributions to precipitation}
\label{subsec:relCon2PcP}

\thispagestyle{empty}
\begin{figure}
	\begin{minipage}[\textheight]{\columnwidth}
		\centering
		\noindent\includegraphics[width = 8cm]{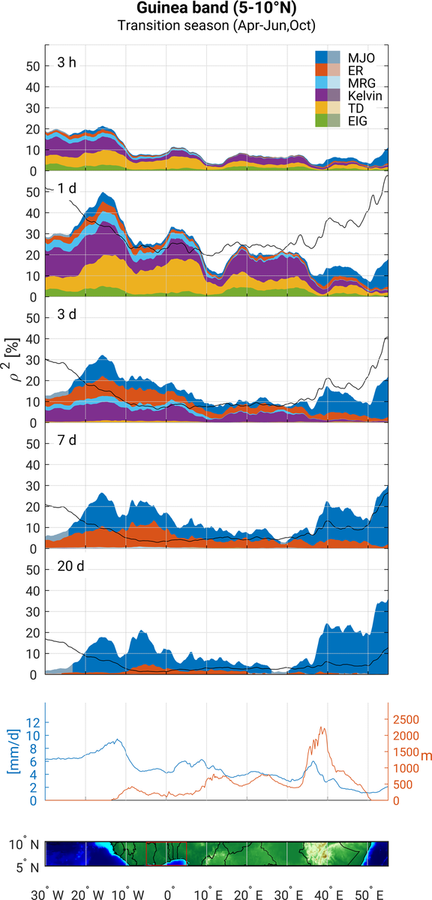}\\
		\caption{Relative importance of tropical wave signals for TRMM precipitation (1998-2016) on different timescales over the Guinean band (\ang{5}-\ang{10}N, bottom panel) during the transition season (April to June and October). Explained variance is estimated by the squared correlations of the wave signal with raw precipitation. \blue{Lines are stacked on top of each other, such that the sum of all lines can be understood as the maximum variance explained by all wave types.} Significant correlations ($p<0.05$) are indicated with saturated colors. The black solid line shows the variance of precipitation at the specific timescale as percentage of the total variance in the raw data (3h). Band-averaged daily precipitation and surface height is given in the second lowest panel.}\label{fig.rhoGuineaTransition}
		\vfill	
	\end{minipage}
\end{figure}

	To understand the relative contribution of the tropical waves to rainfall variability on different timescales, the wave filtered precipitation was correlated with temporally aggregated raw precipitation (Figs.~\ref{fig.rhoGuineaTransition}--\ref{fig.rhoSahelFull}). The squared correlation is calculated as an estimation of the explained variance of raw precipitation by the tropical waves. All waves significantly modulate precipitation over the Sahel and Guinean bands. Using this correlation analysis, we can obtain a more detailed overview over the longitudinally varying modulation strength as well as the relative contribution to rainfall variability on different timescales. Interactions between different waves are not taken into account, so the sum can be seen as the maximum that can be explained by the six tropical waves combined. During the transition season, the Sahel band receives very little rainfall (Fig.~\ref{fig.trmm}) and the influence of tropical waves is comparatively weak (Figs.~\ref{fig.trmmwavevar} and \ref{fig.StationsPDF}b). Thus, a figure for the Sahel band during the transition is not discussed here, but is available in the supplementary material (Fig.~S14). 
	
	For both seasons and both bands, we can observe that the influence of tropical waves depends on the timescale and varies substantially with location and season. As expected, waves in shorter period bands generally dominate the rainfall variability on shorter timescales, whereas the longer timescales are governed by slower waves. \blue{It should be noted that a specific wave modulates precipitation at the accumulation period less or equal to half the wave period. At accumulation times of equal to one period (and longer), the enhanced and suppressed phase cancel each other out and thus no correlation can be seen anymore at this time scale.} 
	
	\begin{figure}
		\begin{minipage}[\textheight]{\columnwidth}
			\centering
			\noindent\includegraphics[width = 8cm]{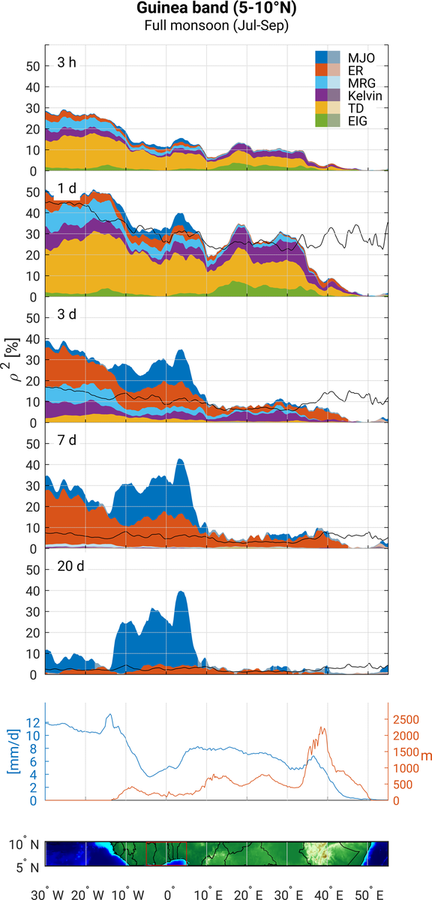}\\
			\caption{Same as Fig.~\ref{fig.rhoGuineaTransition}, but during the full monsoon season (July to September).}\label{fig.rhoGuineaFull}
			\vfill
		\end{minipage}
	\end{figure}
	
	The Guinea band exhibits the strongest modulation on the daily timescale during the transition season (Fig.~\ref{fig.rhoGuineaTransition}). Here, the TD and Kelvin waves have the strongest impact. Between the \blue{Cameroon Line} (\ang{12}E) and Ethiopian Highlands (\ang{38}E), rainfall variability is dominated by Kelvin waves. On the timescale of three days, MJO, ER, and Kelvin waves are approximately equally important. The influence of TDs is not evident any more on this timescale. On the weekly timescale, ER and MJO remain as the only contributors to rainfall variability. On the short subseasonal timescale (20 days), the MJO is the remnant major source of rainfall variability locally explaining 10--\SI{20}{\percent} of rainfall variability over West Africa and up to 20--\SI{30}{\percent} over the Horn of Africa.  A stronger influence of the MJO on Central African precipitation during the spring season was documented by \citet{Berhane.2015}. 
 
	During the full monsoon, the TDs gain importance for rainfall variability in the Guinea band. On the daily timescale, they explain up to \SI{30}{\percent} of the variability over West Africa (Fig.~\ref{fig.rhoGuineaFull}). This is consistent with \citet{Dickinson.2000} and \citet{Lavaysse.2006}, who found numbers of the same order for AEWs. The importance of AEWs on daily precipitation increases westwards, which might be attributed to the barotropic-baroclinic growth they experience on the way towards the Atlantic \citep{Charney.1962,Thorncroft.1994,Hsieh.2005}. The influence of Kelvin waves is reduced compared to the transition season, but still explains up to \SI{10}{\percent} over Central Africa, slightly less than what is recorded by \citet{Mekonnen.2016}. EIG and MRG waves gain importance west of \ang{10}W on the timescale of one to three days. On the synoptic timescale (3--7 days), ER contribute about 10--\SI{20}{\percent} to rainfall variability. The main influence of the MJO on the short subseasonal timescale (7--20 days) lies within the West African monsoon region, explaining locally more than one third of rainfall variability. There is no significant influence of the MJO on Central and East African rainfall.
	
		\begin{figure}
		\begin{minipage}[\textheight]{\columnwidth}
			\centering
			\noindent\includegraphics[width = 8cm]{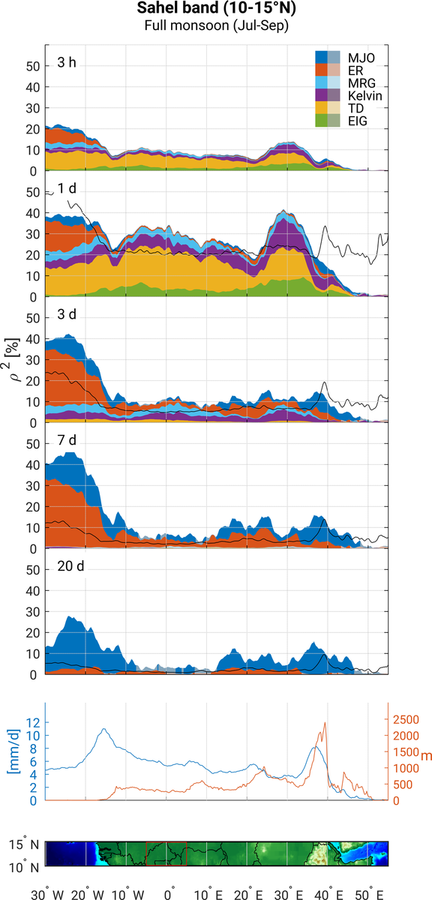}\\
			\caption{Same as Fig.~\ref{fig.rhoGuineaTransition}, but over the Sahelian band (\ang{10}-\ang{15}N) and during the full monsoon season (July to September).}\label{fig.rhoSahelFull}
			\vfill
		\end{minipage}
	\end{figure}
		
	In the Sahel band, TD are the dominant wave on the 3-hourly to daily timescale, explaining up to \SI{20}{\percent} of the rainfall variability during the full monsoon (Fig.~\ref{fig.rhoSahelFull}). A large portion of rainfall variability is concentrated on timescales longer than three days off the West Coast of Africa (black line). From three to seven days, ER waves are a large contributor to rainfall variability in this region, explaining up to one third of rainfall variability. On longer timescales (20 days), only the MJO signal contributes to rainfall variability over the Atlantic Ocean and parts of East Africa. A more pronounced modulation of the MJO in the Guinea band than in the Sahel band was also found by \citet{Gu.2009} and \citet{Pohl.2009}.
	
	The QBZD and the 'Sahel' mode operate on the timescale of 10--25 days. \citet{Mounier.2008} showed that the QBZD has a stationary component and suggested Kelvin waves as triggers for the QBZD. The present correlation analysis suggests that Kelvin waves do not explain rainfall variability on timescales longer than seven days. It has to be noted though, that the correlation analysis cannot take into account cases when Kelvin waves trigger the QBZD, which then affects longer timescales. The 'Sahel mode' is mainly associated with ER waves \citep{Janicot.2010}. ER explain a substantial portion on the weekly timescale; yet, the present analysis suggest, that 'pure' ER waves, as derived from the shallow water equations, are barely relevant on the timescale of 20 days (Figs.~\ref{fig.rhoGuineaTransition}--\ref{fig.rhoSahelFull}). The correlation analysis again might not capture cases when on a longer timescale, MJO events over the Indian ocean trigger single ER and Kelvin waves that meet over the African continent as suggested by \citet{Matthews.2000,Matthews.2004}. The present analysis should be understood as an account of how much variability can be explained by the respective wavenumber-frequency bands, independent of more complex wave interactions.

\subsection{The role of orography}
\label{subsec:orogr}
	\blue{The effect of orography on the propagation of the MJO and the diurnal cycle of precipitation over the Maritime Continent has been widely discussed (e.g. \citealt{Inness.2006,Peatman.2014,Tseng.2017,Sakaeda.2017,Tan.2018}). Over northern tropical Africa, several tropical waves also appear} to be influenced by the presence of orography. Both latidunal bands show a reduced influence of tropical waves over orography (Figs.~\ref{fig.rhoGuineaTransition}-\ref{fig.rhoSahelFull}). In the Guinea band, Kelvin wave and EIG are reduced after the passage of the Guinea Highlands (\ang{12}W), over the \blue{Cameroon Line} (\ang{12}E), and the Ethiopian Highlands (\ang{38}E). The reduced modulation of TD is evident over the Guinea Highlands (\ang{12}W), \blue{Cameroon Line} (\ang{12}E), the Bongo Massif (\ang{25}E), and the Ethiopian Highlands (\ang{38}E). In the Sahel band, the influence of tropical waves is reduced over the Guinea Highlands (\ang{12}W), the Jos Plateau (\ang{8}E), the \blue{Darfur Mountains} (\ang{23}E), and the Ethiopian Highlands (\ang{38}E).  A reduced modulation over the \blue{Cameroon Line} can also be seen in Figs.~\ref{fig.mapER}, \ref{fig.mapAll}, S4, S5 for the ER, Kelvin waves, and TD. In general, the reduced effect of tropical waves over orography is more pronounced on the daily and subdaily timescale, where the waves compete directly with diurnal circulations, and is of the order of \SI{50}{\percent}. The effect of orography, and the Ethiopian Highlands in particular, on Kelvin waves has been documented by \citet{Matthews.2000}. Kelvin waves encountering orography can be deflected away from the equator similar to coastal Kelvin waves in the ocean \citep{Gill.1977,Hsu.2005}. Additionally the orographic lifting and frictionally induced lower-tropospheric convergence create a phase shift \citep{Hsu.2005}. Figure S4 also suggests a that convection over the \blue{Cameroon Line} is shifted by one phase. AEW preferably form in the lee of mountain ranges, where vorticity is increased \citep{Mozer.1996}, and in consequence of mesoscale convective systems that have been triggered over the mountains \citep{Mekonnen.2006,Thorncroft.2008}. The present study illustrates how orography significantly reduces the modulation of tropical waves \blue{over northern tropical Africa}.

\subsection{Wave interactions}

\begin{figure*}
	\centering
	\noindent\includegraphics[height=11.4cm]{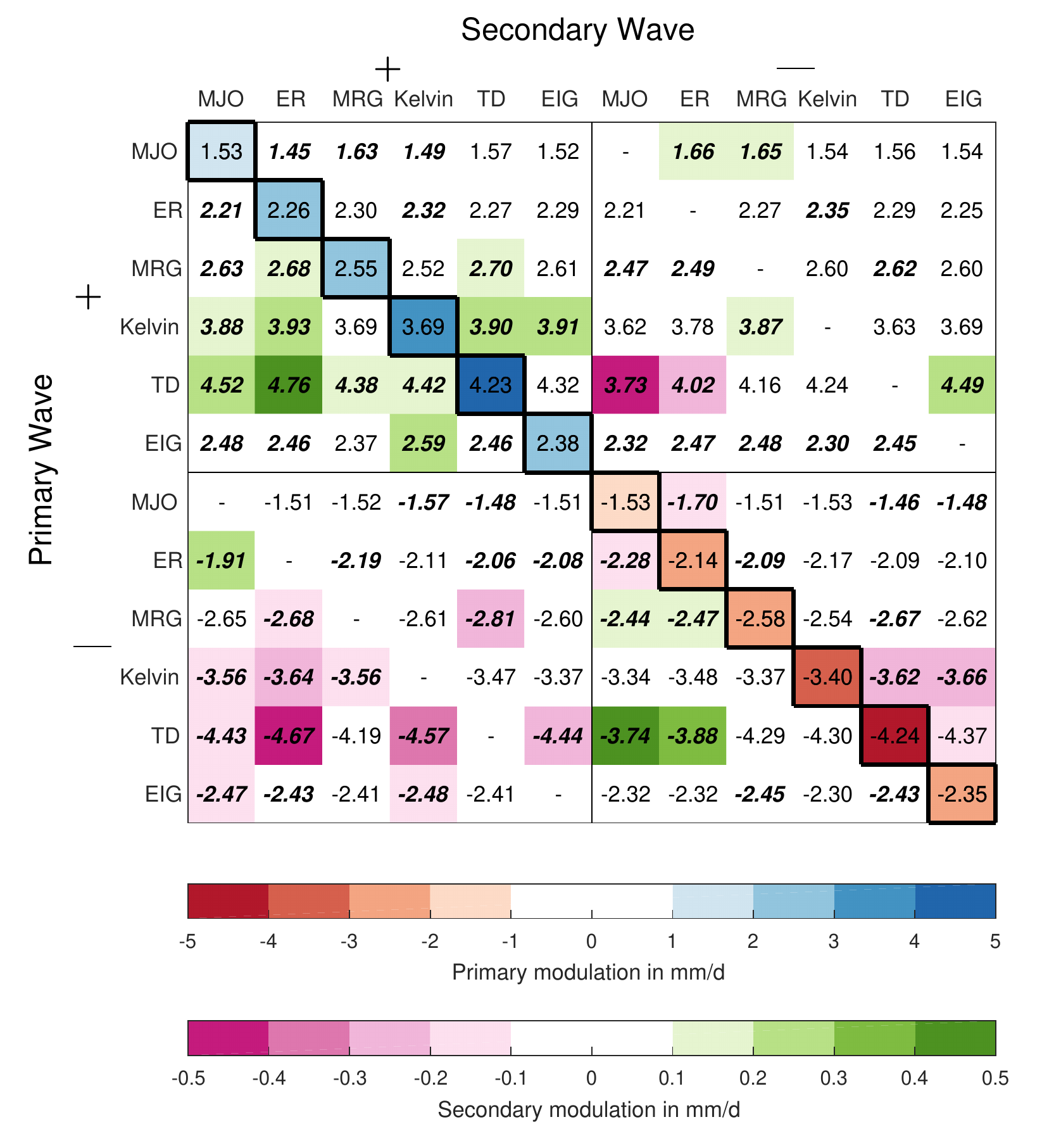}\\
	\caption{Interaction of different wave types in the Guinea Coast box. The mean or primary modulation of each wave measured as the mean wave filtered TRMM precipitation is given on the diagonal. Cases with a wave activity exceeding $\pm1\sigma$ are defined as wet (-) and dry (+) phases. The superposition of a secondary wave is measured as the primary wave signal conditional on the occurrence of a secondary wave. Significant deviations from mean modulation are marked bold and italics ($p<\SI{5}{\percent}$). See section~\ref{subsec:interaction} for more details.}
	\label{fig.interaction}
\end{figure*}

	Finally, interactions between different types of tropical waves over the two case study regions Guinea Coast and West Sahel are examined. Only the results for the Guinea Coast are presented here because the results are very similar for both regions. Figure~\ref{fig.interaction} shows the interaction between a primary wave and a secondary wave for all analyzed wave types in the Guinea Coast. The primary modulation in the Guinea Coast (shown on the diagonal) reflects the pattern already seen in Fig.~\ref{fig.trmmwavevar}. The modulation by a second wave is seen in the entries off the diagonal of the matrix. The magnitude of the secondary modulation depends on the magnitude of the primary modulation and is strongest for TD. In many cases, waves superimpose constructively. The wave activity is generally further enhanced when a primary wave in its positive phase is modulated by a second wave in a positive phase (upper left quarter). Similarly, primary waves in a negative phase modulated by a negative second wave result in a further suppression (lower right quarter). Two interesting exceptions of this wave superpositions can be seen for TD and MRG. Here, a modulation of the negative primary wave by a negative ER or MJO result in a more moderate modulation. Similarly, non-linear wave superpositions are evident when a positive wave is superimposed with a negative wave (lower left and upper right quarters). \blue{This means that both low-frequency waves in their wet phases amplify the high-frequency waves and suppress them in the dry phase.} The analysis for the West Sahel reveals similar patterns (Fig.~S15, supplementary material). A dominant difference is a missing modulation of TD by the dry phases of ER and MJO waves.
	
	According to theory, equatorial waves behave as linear or independent solutions of the shallow-water system \citep{Matsuno.1966}. This is, of course, an idealized view; the basic analysis of wave interactions presented here suggest that also the dry phase of TD during the passage of the suppressed phase of MJO and ER events are significantly weaker. The superposition of AEWs and MJO events was also analyzed by \citet{Ventrice.2011}. Decreased AEW activity was found when the MJO is active over the western Pacific and Atlantic, AEW activity was enhanced in phases where the MJO is active over Africa and the Indian Ocean. The superposition of MJO, Kelvin and ER waves over Vietnam was analyzed in detail by \citet{vanderLinden.2016}. The presented results suggest complex superpositions that require to be analyzed in more detail.


\section{Conclusion}  \label{sec:Summary}

The impact of tropical waves on rainfall variability over northern tropical Africa has long been known but the relative influence of the different tropical wave types has never been quantified for this region. As previous studies applied different methodologies and datasets, a systematic comparison of the results has been difficult. This study closes this gap and gives a first comprehensive and systematic account of the influence of all major tropical wave types for daily to intraseasonal rainfall variability over northern tropical Africa. A consistent method was applied to all waves and modulation intensities were quantified comparing two satellite datasets and data from a dense rain gauge network. The new quantification of rainfall anomalies using quantile anomalies made it possible to compare the influence on precipitation in different climatic zones and revealed influences deep into the subtropics. All analyzed waves were found to significantly modulate rainfall variability in the monsoon system on different temporal scales. The main findings are:

\begin{itemize}
	\item Equatorial waves contribute most to rainfall variability over northern tropical Africa in the area of the seasonal rainfall maximum. TD and Kelvin waves explain the overall highest variability \blue{followed equally by MRG and ER waves}.
	
	\item \blue{Precipitation patterns of tropical waves are mainly confined to the tropics. Notably, the influence of MJO and ER wave reaches deep into the subtropics and their modulation patterns resemble tropical plumes \citep{Knippertz.2005,Frohlich.2013}.} The different datasets show comparable modulation intensities varying from less than 2 to above \SI{7}{\mm\per\day} depending on the wave type.
	
	\item Tropical waves modulate precipitation on different timescales. The influence varies with location and season. On the 3-hourly to daily timescale, TD and Kelvin waves are the dominant wave types. On longer timescales (7--20d), only MJO and ER remain as modulating factors for rainfall variability. For the first time, this study analyzed the influence of EIG on northern Africa. Over Central and East Africa, EIG waves explain 5--\SI{10}{\percent} of rainfall variability on the daily timescale. The rainfall modulation by several tropical waves is reduced over orography.
	
	\item Tropical waves superimpose and interact with each other. \blue{During their wet phases, the low-frequency waves (MJO and ER) amplify the high-frequency TD and MRG waves and suppress them in their dry phase.}
	
\end{itemize}

Several open questions remain. The present study only gives an observational account of tropical waves. It remains unclear how the waves influence the dynamic and thermodynamic environment as well as how interactions with the West African monsoon system work. A follow-up paper is in preparation, which will deal with these open questions in order to get a better understanding how the different waves differ in the way they modulate the dynamics and thermodynamics of the West African monsoon.

This study emphasizes the need of an adequate representation of tropical waves in NWP models for weather prediction over northern tropical Africa. Further research is needed to assess and improve the prediction of the correct phase and intensity of the waves. Additionally, the statistical relationships found here indicate that, besides conventional NWP models, statistical forecast models including tropical waves may provide useful predictions, bearing in mind that so far global NWP models fail to reliably predict precipitation over this region \citep{Vogel.2018}. First promising results have already been achieved by the authors and are currently investigated in more detail.

Finally, the results stress the importance of tropical waves for operational forecasting of precipitation on different timescales for this region. The presented new visualization showing the contribution of tropical waves on precipitation variability on different timescales is useful for weather forecasters to determine the relevance of different waves for their region depending on the analyzed timescale. Further research and collaboration with national and pan-African weather agencies should be encouraged to foster the knowledge transfer to local societies.
\\
\bigskip

%
\textbf{Acknowledgment}

\footnotesize
The research leading to these results has been accomplished within project C2 ``Prediction of wet and dry periods of the West African Monsoon'' of the Transregional Collaborative Research Center SFB / TRR 165 ``Waves to Weather'' funded by the German Science Foundation (DFG). \blue{The authors thank George Kiladis and two anonymous reviewers whose comments helped to improve and clarify this manuscript.} We also thank Tilmann Gneiting for discussions and comments on draft versions of the paper. The authors also thank various colleagues and weather services that have over the years contributed to the enrichment of the KASS-D database; special thanks go to Robert Redl for creating the underlying software. The authors also thank Roderick van der Linden for providing code that was further developed to filter the waves and create the composite plots. Finally, we thank Carl Schreck, III for providing the filtering function \textit{kf\textunderscore filter}.

%

\vspace{5mm}

\bibliography{references}

\onecolumn

\center \large \textbf{Supplementary Material}
\renewcommand{\thefigure}{S\arabic{figure}}

\setcounter{figure}{0} 

\begin{figure}[h!]
	\centering
	\noindent\includegraphics[width = 0.95 \textwidth ]{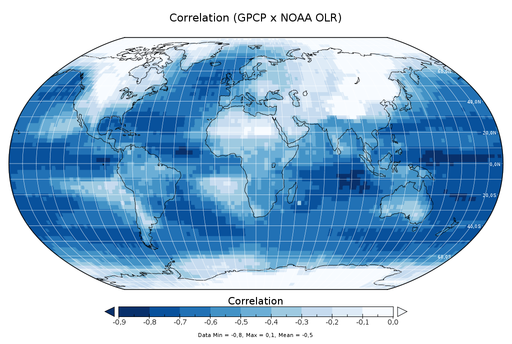}\\
	\caption{Correlation of daily precipitation from the Global Precipitation Climatology Project (GPCP) and NOAA OLR. Note the low correlations in the extratropics, the subtropics, and over West Africa}\label{fig.GPCPOLR}
\end{figure}

\begin{figure*}[p]
	\centering
	\noindent\includegraphics[height = 0.9 \textheight ]{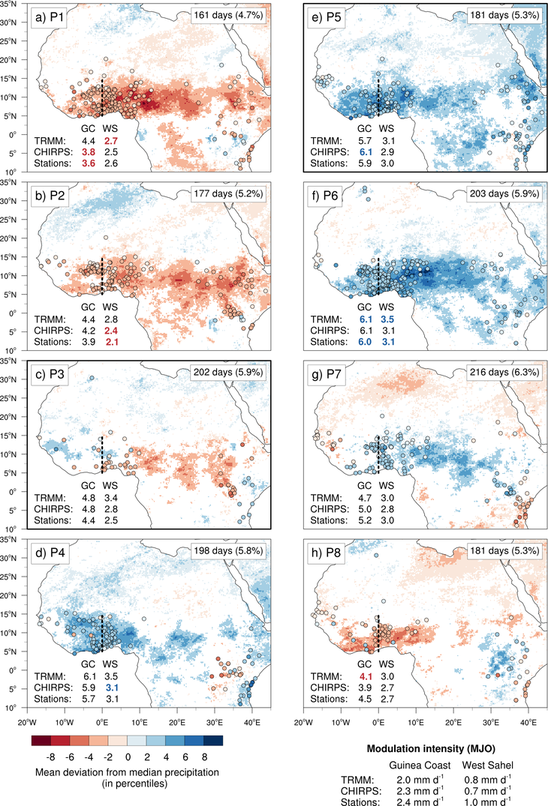}\\
	\caption{Same as Fig.~5 in the main paper, but for Madden-Julian Oscillation.}
\end{figure*}

\begin{figure*}[p]
	\centering
	\noindent\includegraphics[height = 0.95 \textheight ]{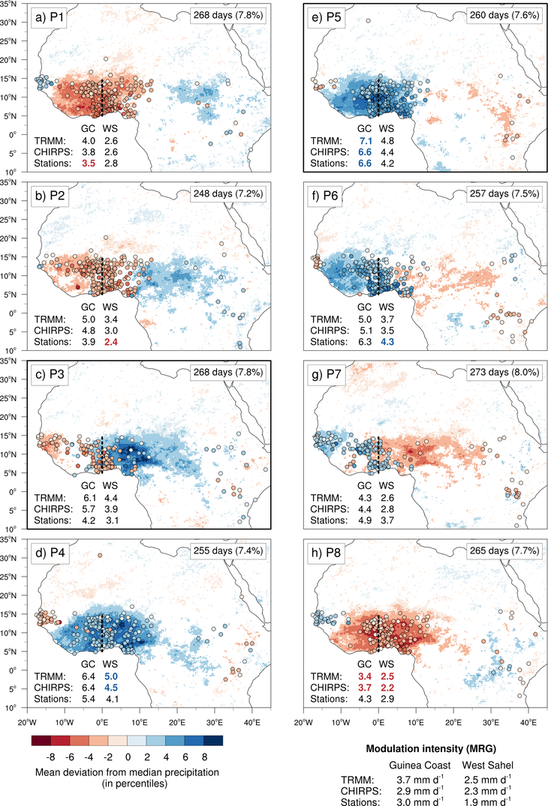}\\
	\caption{Same as Fig.~5 in the main paper, but for mixed Rossby gravity waves.}
\end{figure*}

\begin{figure*}[p]
	\centering
	\noindent\includegraphics[height = 0.95 \textheight ]{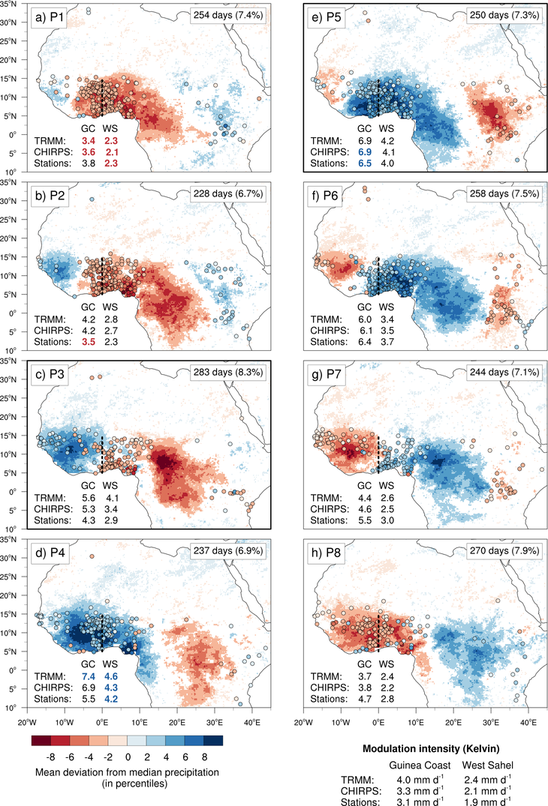}\\
	\caption{Same as Fig.~5 in the main paper, but for Kelvin waves.}
\end{figure*}

\begin{figure*}[p]
	\centering
	\noindent\includegraphics[height = 0.95 \textheight ]{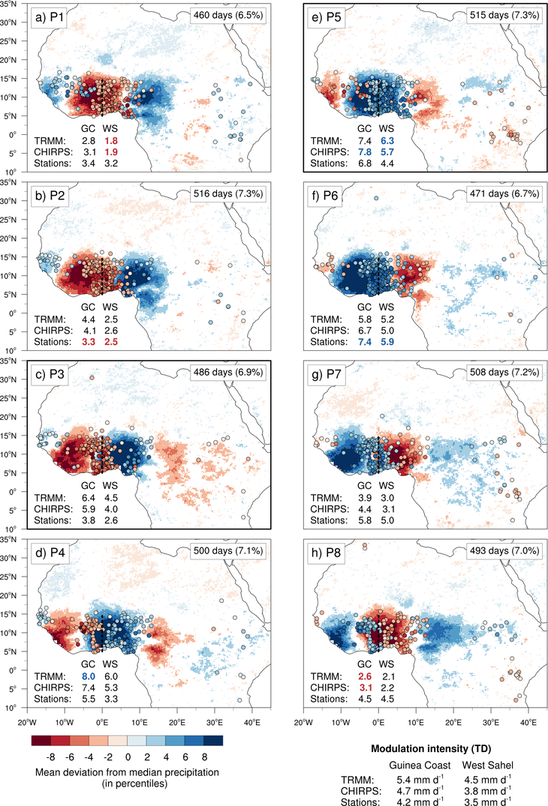}\\
	\caption{Same as Fig.~5, but for tropical disturbances (mostly African Easterly Waves).}
\end{figure*}

\begin{figure*}[p]
	\centering
	\noindent\includegraphics[height = 0.95 \textheight ]{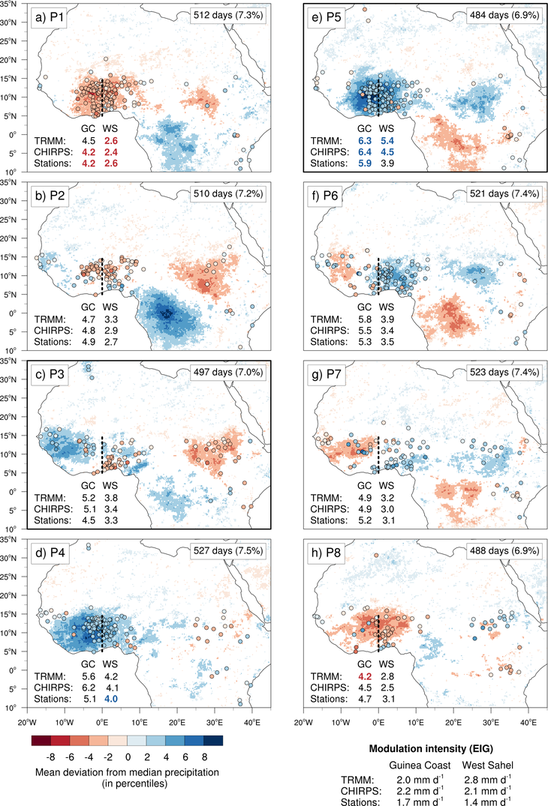}\\
	\caption{Same as Fig.~5 in the main paper, but for eastward inertio-gravity waves.}
\end{figure*}

\begin{figure*}[p]
	\centering
	\noindent\includegraphics[height = 0.95 \textheight ]{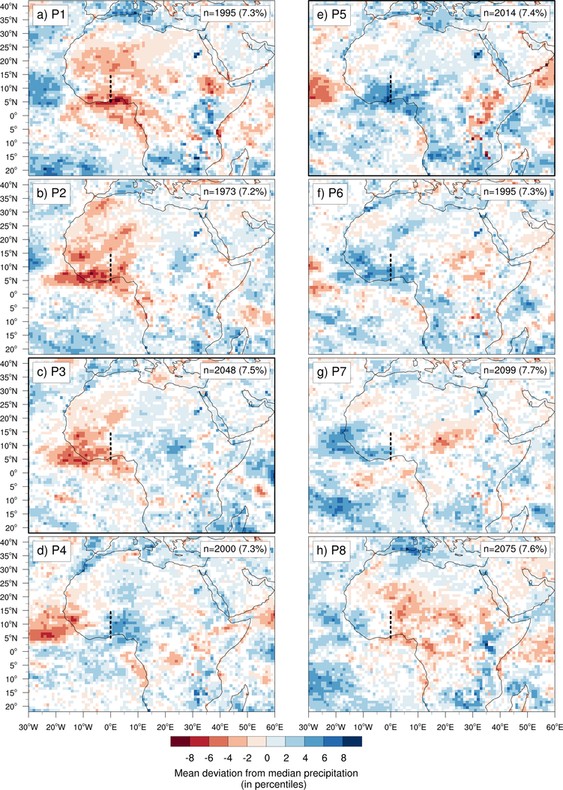}\\
	\caption{Same as Fig.~5 in the main paper showing a composite of the equatorial Rossby wave, but filtering based on 3h-TRMM precipitation instead of NOAA OLR and showing TRMM rainfall anomalies. TRMM has been spatially aggregated to a \ang{1}x\ang{1}-grid and temporally aggregated to daily values before the calculation of quantiles at each grid point. This helps to obtain smoother results, because random errors are reduced by increasing the sample size. The filtering period ranges from 1998--2013.}\label{fig.mapER_TRMM}
\end{figure*}

\begin{figure*}[p]
	\centering
	\noindent\includegraphics[height = 0.95 \textheight ]{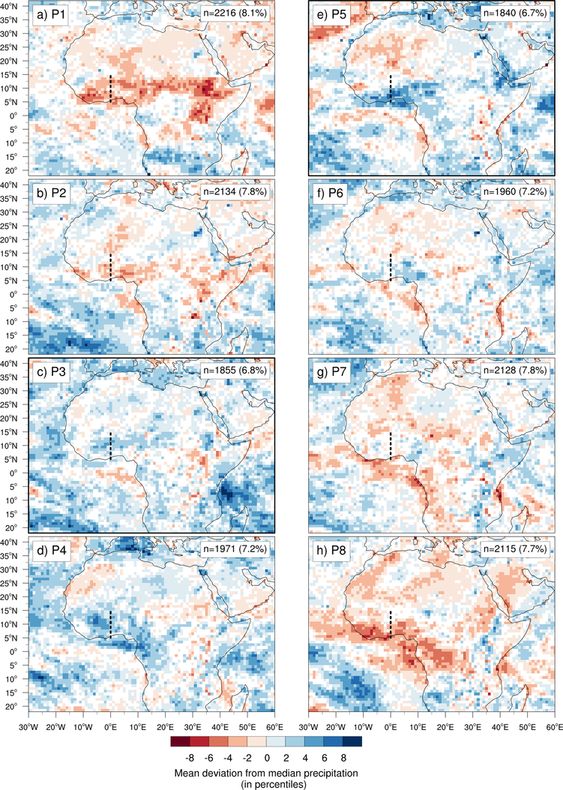}\\
	\caption{Same as Fig.~\ref{fig.mapER_TRMM}, but for Madden-Julian Oscillation.}
\end{figure*}

\begin{figure*}[p]
	\centering
	\noindent\includegraphics[height = 0.95 \textheight ]{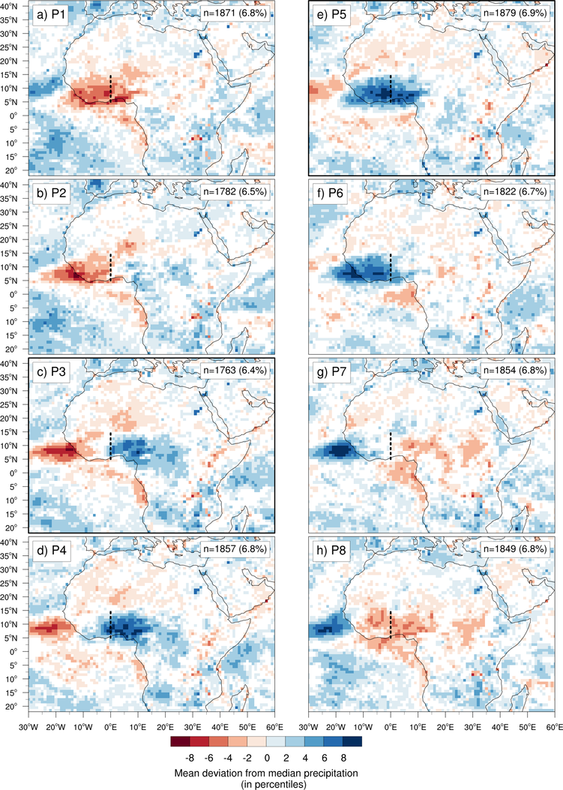}\\
	\caption{Same as Fig.~\ref{fig.mapER_TRMM}, but for mixed Rossby gravity waves.}
\end{figure*}

\begin{figure*}[p]
	\centering
	\noindent\includegraphics[height = 0.95 \textheight ]{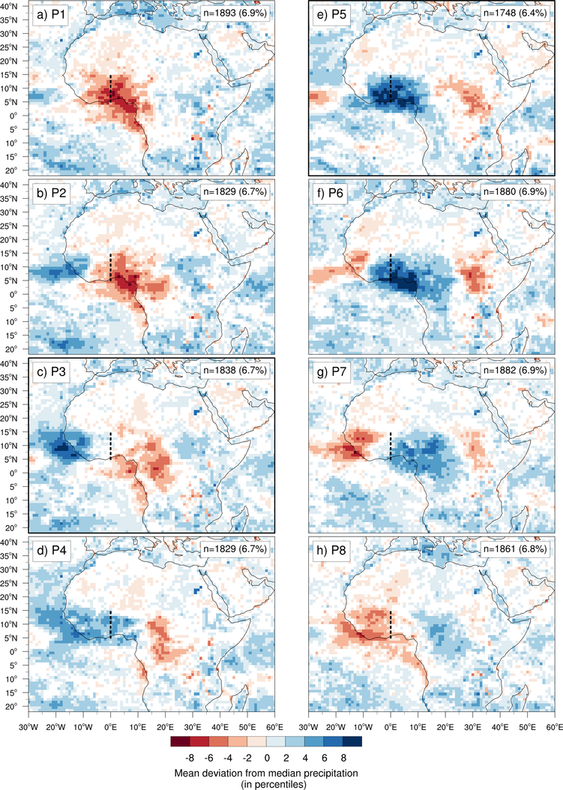}\\
	\caption{Same as Fig.~\ref{fig.mapER_TRMM}, but for Kelvin waves.}
\end{figure*}

\begin{figure*}[p]
	\centering
	\noindent\includegraphics[height = 0.95 \textheight ]{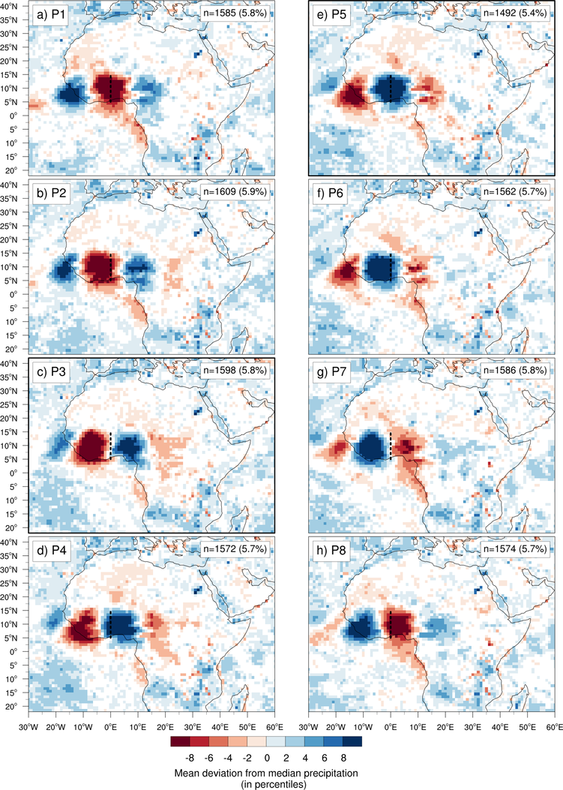}\\
	\caption{Same as Fig.~\ref{fig.mapER_TRMM}, but for tropical disturbances (mostly African Easterly Waves).}
\end{figure*}

\begin{figure*}[p]
	\centering
	\noindent\includegraphics[height = 0.95 \textheight ]{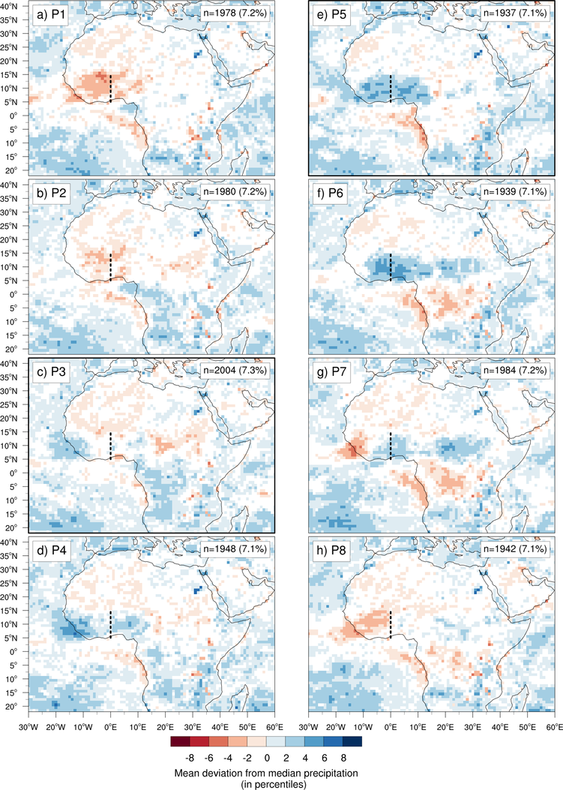}\\
	\caption{Same as Fig.~\ref{fig.mapER_TRMM}, but for eastward inertio-gravity waves.}
\end{figure*}

\begin{figure*}[p]
	\centering
	\noindent\includegraphics[width = 0.95 \textwidth ]{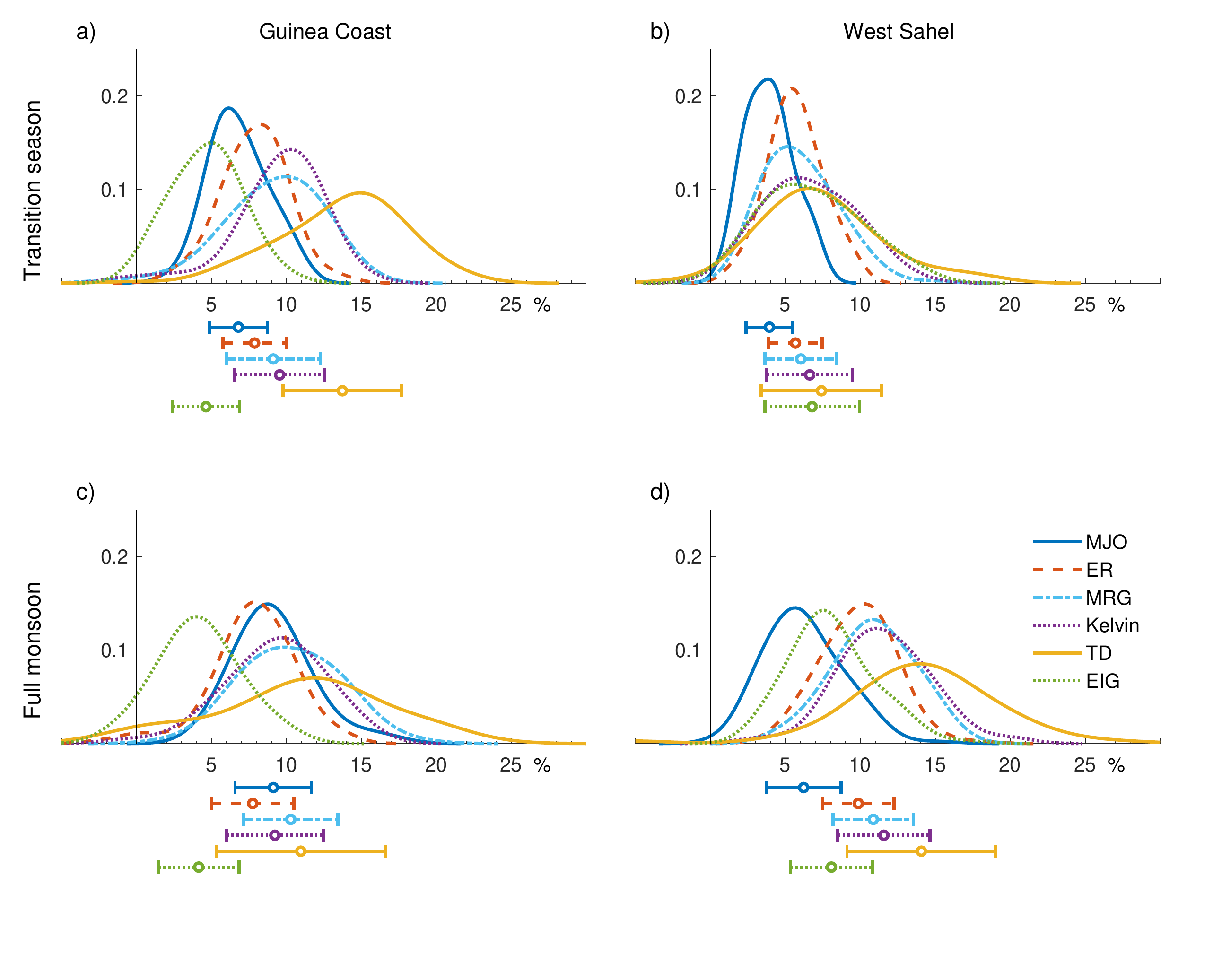}\\
	\caption{Same as Fig.~8 in the main paper, but measured in quantile anomalies (for more details see subsection e in methods section).}
\end{figure*}

\begin{figure*}[p]
	\centering
	\noindent\includegraphics[height = 0.95 \textheight]{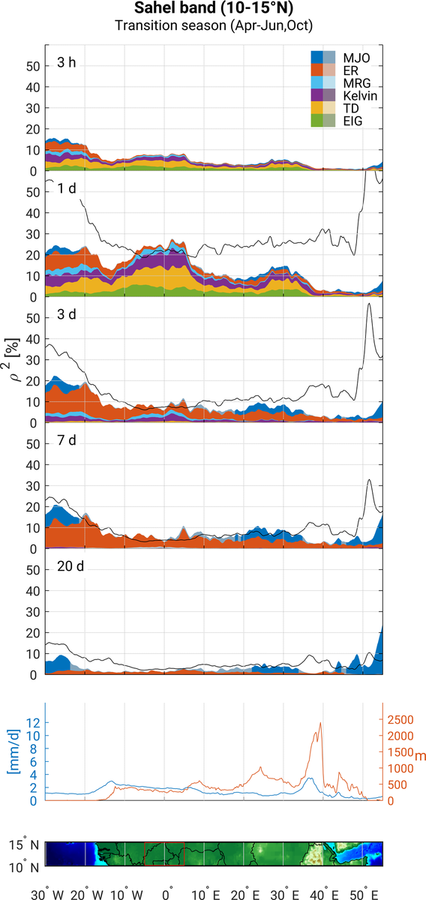}\\
	\caption{Same as Fig.9 in the main paper, but over the Sahelian band (\ang{10}-\ang{15}N) and during the transition season (April to June and October).}
	\vfill
\end{figure*}

\begin{figure*}[p]
	\centering
	\noindent\includegraphics[height=11.4cm]{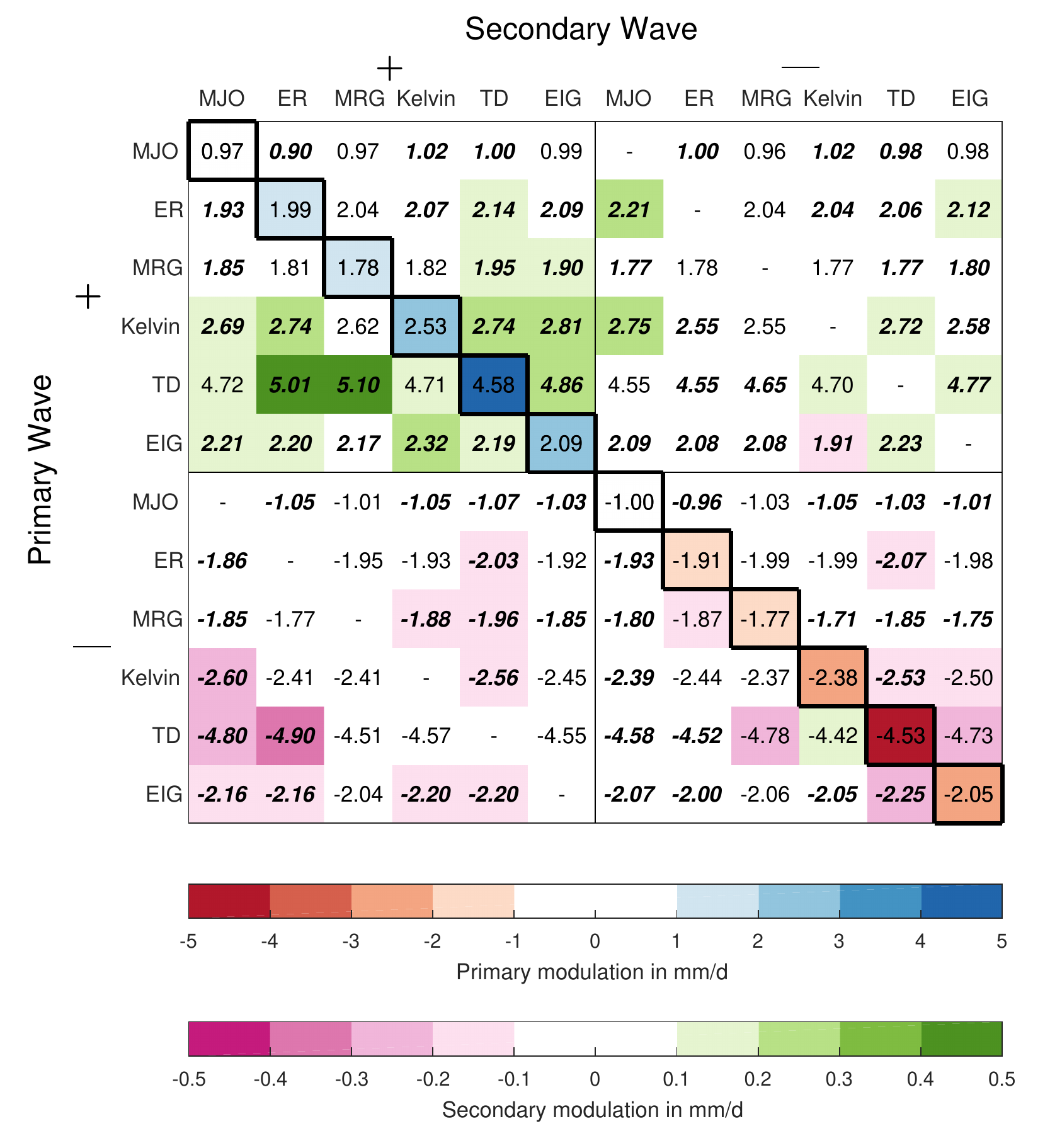}\\
	\caption{Same as Fig.~12 in the main paper, but for the West Sahel box (\ang{5}W--\ang{5}E, \ang{10}--\ang{15}N).}
\end{figure*}

\end{document}